\documentclass[runningheads]{llncs}
\usepackage[misc]{ifsym} 
\usepackage{graphicx}
\usepackage{amsmath,amssymb,amsfonts}
\usepackage[linesnumbered,algoruled,boxed,lined,noend]{algorithm2e}
\SetAlFnt{\small\sf}
\usepackage{epstopdf}
\usepackage{subfigure}
\usepackage{textcomp}
\usepackage{dsfont}
\usepackage{xcolor}
\usepackage{todonotes}

\begin{document}
\title{PrivSketch: A Private Sketch-based Frequency Estimation Protocol for Data Streams}%\thanks{Supported by organization x.}}
\titlerunning{PrivSketch: A Private Sketch-based Frequency Estimation Protocol}
\author{Ying Li\inst{1,2} \and
Xiaodong Lee\inst{1,2}\textsuperscript{(\Letter)} \and
Botao Peng \inst{1}\textsuperscript{(\Letter)} \and
Themis Palpanas \inst{3} \and
Jingan Xue \inst{4}
\institute{Institute of Computing Technology, Chinese Academy of Sciences \and 
University of Chinese Academy of Sciences \and
LIPADE, Universit{\'e} Paris Cit{\'e} French University Institute (IUF) \and
Huawei Technologies
}
}
\authorrunning{Ying Li et al.}
\maketitle              % typeset the header of the contribution

\begin{abstract}
		Local differential privacy (LDP) has recently become a popular privacy-preserving data collection technique protecting users' privacy.% without a trusted collector. 
		The main problem of data stream collection under LDP is the poor utility due to multi-item collection from a very large domain.  
		This paper proposes PrivSketch, a high-utility frequency estimation protocol taking advantage of sketches, suitable for private data stream collection.
		Combining the proposed background information and a decode-first collection-side workflow, PrivSketch improves the utility by reducing the errors introduced by the sketching algorithm and the privacy budget utilization when collecting multiple items. 
		We analytically prove the superior accuracy and privacy characteristics of PrivSketch, and also evaluate them experimentally. 
		Our evaluation, with several diverse synthetic and real datasets, demonstrates that PrivSketch is 1-3 orders of magnitude %38x-2098x
		better than the competitors in terms of utility in both frequency estimation and frequent item estimation, while being up to $\sim$100x faster. % in execution time.
  This paper was published in DEXA 2023.

%\keywords{LDP  \and Sketch \and Frequency estimation.}
\end{abstract}

\section{Introduction}

\textbf{Motivation.} 
Collecting user data, often in the form of a data stream, in order to analyze them and provide some services has become a common practice.
However, data collection may expose user information,% and harm the user privacy, 
which is a major concern.
Local Differential Privacy (LDP) is popular to protect individual privacy during data collection and has been widely used in technology companies (such as Apple, Google, Microsoft). 
It perturbs data locally before sending them to the collector and enables the collector to obtain approximate statistics on the perturbed data, to avoid the risk of disclosing user privacy. 
A parameter $\epsilon$ is used to quantify the amount of perturbation, which determines the degree of privacy protection and the utility of the privacy-preserving algorithm.

\noindent \textbf{Utility Problem.} 
Although several studies have focused on the frequency estimation problem under LDP, they do not perform well when used in a data stream context, due to following reasons. 
First, existing solutions consider a unified size for the data items generated by different users (i.e., data length) that is based on unrealistic assumptions (assume only one item in a collection interval)~\cite{Ding2017,wang2021continuous}, or a predefined/estimated unified size $L$ for each collection (padding and sampling)~\cite{LDPMiner,ShaoweiWang2018,Wang2020}, in both cases hurting utility.
Second, the large domains of several data streams (e.g., URLs, and IP) lead to excessive computation and communication costs, as well as significant perturbation errors (some of the existing literature on frequency estimation~\cite{ShaoweiWang2018,Wang2020} is only applicable to small cardinality domains).
Sketching is widely used in streaming data processing for compressing sparse data from a large domain (e.g., when a tiny percentage of webpages are accessed by any individual user).
The uniform size of sketches makes it possible to unify the data length of different users without extra padding and sampling~\cite{LDPMiner}, and leads to efficient storage. % solutions for large domain data.
The sketching has been combined with LDP in the Private Count-Mean Sketch (PCMS) algorithm~\cite{Apple2017} proposed by Apple.
However, it operates at the granularity of single items, which hurts performance. 
When considering extending it to multi-item collections, the following problems emerge. 
(\romannumeral1) The error introduced by sketching algorithms is not considered. 
Aggregating sketches from users directly is equivalent to encoding all data into one sketch, leading to increasing errors of collisions, i.e., data hashed in the same position.
(\romannumeral2) To maintain user-level privacy, allocating the privacy budget for each counter in the sketch is required, resulting in substantial inaccuracies and poor utility. 

\noindent \textbf{Our solution.} 
%To meet the above challenges, we 
We propose PrivSketch, a high-utility privacy-preserving sketch-based frequency estimation protocol that leads to lower errors when compared to existing solutions. 
PrivSketch proposes an innovative LDP collector-side workflow that decodes the perturbed sketch before aggregating and calibrating it,
which avoids the error introduced by the collisions when aggregating all perturbed sketches in the traditional decode-after workflow.
In addition, PrivSketch utilizes the ordering matrix extracted from the original sketch, 
which enables the collector to obtain the minimum index information, 
while ensuring the privacy of each user's sketch (cf. proof in Section~\ref{PrivacyAnalysis}). 
This effectively reduces the minimum calculation error caused by disturbance and is the first attempt to improve utility using background information.
Furthermore, PrivSketch uses the sampling technique to improve the utilization of the information encoded by the sketch, transmitting relatively accurate information with a limited privacy budget. 
Thus, it reduces the error caused by the uniform allocation of the privacy budget when encoding multiple items.

\textbf{Contributions.} 
Our contributions are summarized below.

%\begin{itemize}
%\item 
$\bullet$
We propose a novel LDP protocol, PrivSketch, %based on sketching, 
that is suitable for frequency estimation in data streams %network traffic collection, 
where multi-item encoding is needed. 
It is the first sketch-based privacy-preserving protocol that considers the errors introduced by the sketches with a novel decode-first workflow.
%PrivSketch decodes before aggregating. %based on the analysis of the errors introduced by the sketching algorithm. 
%At the same time, 
 It employs background information to reduce the minimum value calculation errors of sketches and utilizes a sampling technique to improve the privacy budget utilization. %achieving higher utility than existing algorithms.
	
%\item 
$\bullet$
We prove (cf. Section~\ref{PrivacyAnalysis}) that the ordering matrix as background information does not expose the original value of each counter in the sketch, which meets the privacy needs of users. %under $\epsilon$-LDP. 
We introduce a new definition of the indistinguishable input set, where the collector cannot distinguish any two values.% in this set. 
We observe that the utility of LDP algorithms can be improved with appropriate additional background information
%that affects the size of the indistinguishable input set
, but does not harm the users' privacy.  
	
%\item 
$\bullet$
We evaluate our approach on both synthetic and real datasets. 
We compare it with extensions and variants of existing algorithms, including the multi-item encoding extension and its min-variant algorithm. 
The utility of the protocol proposed in this paper is 1-3 orders of magnitude more accurate than existing algorithms, and up to $\sim$100x faster. 

\section{Background and preliminaries}

\textbf{Local Differential Privacy (LDP).} 
Differential privacy (DP)~\cite{dwork2006calibrating} is a technology with quantified privacy protection, but relies on a trustworthy third-party collector. 
%In the real world, the collector sometimes are attacked resulting in data breaches.
To remove the trust in the collector, LDP~\cite{duchi2013local} was proposed  %does not rely on a trusted collector. 
%LDP guarantees users' privacy locally, so 
where original data are only accessible by users, and the collector only receives the perturbed data.  
A mechanism $\mathcal{M}$ satisfying LDP can be defined as follows. %with a privacy budget $\epsilon > 0$ as follows.
\begin{definition}[$\epsilon$-Local Differential Privacy~\cite{duchi2013local}]
A randomized algorithm $\mathcal{M}$ satisfies $\epsilon$-local differential privacy ($\epsilon>0$), if and only if for any two input tuples $x, x' \in \mathcal{D}$ and output $y$, then %the following inequality always holds.
$\frac{\Pr[\mathcal{M}(x)=y]}{\Pr[\mathcal{M}(x')=y]}\le e^{\epsilon}$. 
\end{definition}
\noindent Thus, a smaller $\epsilon$ means large perturbation and more indistinguishable, but lower utility. There is an important property of LDP: 
\begin{theorem}[Sequential Composition Mechanism~\cite{mcsherry2009privacy}] \label{theorem:sequential-ldp}
	Assume a randomized algorithm $\mathcal{M}$ consists of a sequence of randomized algorithms $\mathcal{M}_i(1\le i\le t)$. When for each $i, \mathcal{M}_i$ satisfies $\epsilon_i$-LDP, $\mathcal{M}$ satisfies $\sum_{i=1}^t\epsilon_i$-LDP.
\end{theorem}
\noindent \textbf{Randomized Response Mechanism (RR)~\cite{warner1965randomized,RAPPOR}.} This fundamental LDP mechanism achieves plausible deniability by allowing users not to give the original value. 
Specifically, for binary values, users answer the original value with probability $p$, and the opposite value with probability $q\!=\!1\!-\!p$. 
To achieve $\epsilon$-LDP, the worst case is
	$\frac{\max\Pr[\mathcal{M}(x)=y]}{\min\Pr[\mathcal{M}(x')=y]} = \frac{p}{1-p}=e^{\epsilon}$, 
therefore $p\!=\!\frac{e^\epsilon}{1+e^\epsilon}$. 
Denote $\Pr[x\!=\!1]$ the percentage of $x\!=\!1$. 
For the collector, %the following equations hold:
$\Pr[y\!=\!1] =p\Pr[x\!=\!1] + (1-p)(1-\Pr[x\!=\!1])$ and $\Pr[y\!=\!0] =p(1-\Pr[x\!=\!1])+(1-p)\Pr[x\!=\!1]$.
$\Pr[y=1]$ and $\Pr[y=0]$ represent the probability of the output $y$ taking the value of $1$ and $0$, respectively, 
which can used to obtain the unbiased estimation of $\Pr[x=1]$ and $\Pr[x=0]$.

\noindent \textbf{Count-Min Sketch (CMS).}
A common approach to compress data from a large domain is the sketching algorithm. 
The Count-Min Sketch~\cite{cormode2005improved} is one of the most popular sketching algorithms due to its efficiency.
The sketching uses a matrix $X$ consisting of $K\times M$ counters, bound to $K$ hash functions $H_1,H_2,\dots,H_K: \{1,\dots, d\} \mapsto \{1, \dots, M\}$. 
It consists of two phases: (\romannumeral1) update, where $K$ hash functions are used to hash the updated item $x$, and then the corresponding counters are updated, i.e. $X_{k,H_k(x)} = X_{k,H_k(x)} + 1, \forall 1\le k\le K$; 
(\romannumeral2) query, where item $x$'s count $c(x)$ is estimated, denoted by $\widetilde{c}(x)$, based on the corresponding counters in the sketch, i.e. $\min_{1\le k\le K}X_{k,H_k(x)}$~\cite{cormode2005improved}.

\noindent \textbf{Private Count-Mean Sketch (PCMS-Mean)~\cite{Apple2017}.}
\label{AppleCMS} 
PCMS-Mean estimates frequency under LDP, where the user perturbs data before sending them to the collector. 
Specifically, for item $x$, each user chooses a hash function $H_k$ and updates $X_{k, H_k(x)}=1$ (other positions keep as $-1$), then, perturbs $X_k$ using RR and sends the perturbed result $\hat{X}_k$ to the collector. 
The collector constructs a matrix of size $K\times M$ where each row is the sum of the perturbed rows indexed by $k$, and estimates the frequency by averaging the sum of $k$ counters corresponding to $K$ hash functions. 
The algorithm assumes that each user generates only one item. 
Thus, for any two rows from different users $X_k$ and $X'_{k'}$, at most two positions can be different. 
To protect these two positions under privacy budget $\epsilon$, the parameter $p$ in 
RR is set to $\frac{e^{\epsilon/2}}{1+e^{\epsilon/2}}$ (cf. Theorem~\ref{theorem:sequential-ldp}). 

When extending PCMS-Mean to encode multiple items, the number of different positions in any two rows from different users is up to $M$ due to unlimited items of each user.
Thus, to protect the privacy of each position, the parameter $p$ is set to $\frac{e^{\epsilon/M}}{1+e^{\epsilon/M}}$. 
This naive solution works poorly when $M$ is large. 
The irrational allocation of $\epsilon$ is one of the reasons. 
In addition, the error introduced by the sketching algorithm is also non-negligible. 
The estimation error of different sketching algorithms varies. The error 
of the Count-Min Sketch is smaller than that of the Count-Mean Sketch~\cite{PureLDP};
hence, we use the Count-Min Sketch.

\noindent \textbf{Problem Definition} \label{problem} This paper studies the frequency estimation problem under LDP for data streams, where data are generated from a very large domain. 
There is an untrusted collector and a set of $n$ users represented by $U=\{U_1,U_2,\ldots,U_n\}$. 
Each user, $U_i$, has a set of items of length $L^{(i)}(L^{(i)}\ge 0)$, which is denoted by $S^{(i)}=\{S_1^{(i)},S_2^{(i)},\ldots\}, |S^{(i)}|=L^{(i)}$. 
Each item $S_\ell^{(i)} (0\le\ell\le L^{(i)})$ is discrete value and drawn from a large domain $\mathcal{D}$ of size $|\mathcal{D}|=d$, that is, $S_\ell^{(i)} \in \mathcal{D}$. 
In this paper, we focus on estimating the frequency of each item from $\mathcal{D}$, that represents the proportion of users who possess the item. 
Formally, the frequency for each value $x \in \mathcal{D}$ is defined as:
% {\small
% \begin{equation} \label{eqn:freq-definition}
	$f(x) = \frac{|\{i | \exists \ell, 0\le\ell\le L^{(i)}, S_\ell^{(i)}=x\}|}{n}$.
% \end{equation}
% }
\section{PrivSketch Solution}
\label{overview}

PrivSketch is a LDP protocol based on CMS to solve the frequency estimation problem in data stream collection. 
PrivSketch uses a novel collector-side workflow (cf. Section~\ref{orderingmatrix}) and the ordering matrix (cf. Section~\ref{orderingmatrix}) to reduce errors introduced by sketches. 
PrivSketch also uses a sampling technique to increase the information utilization in sketches under a limited privacy budget. 

\begin{figure*}[tb]
	\centering
	\includegraphics[width=\textwidth]{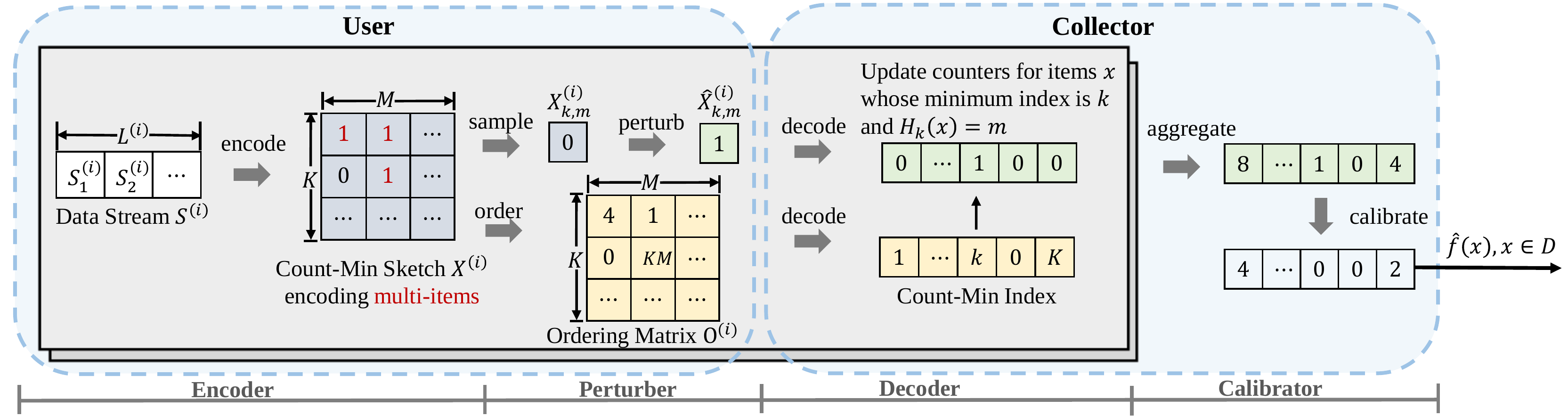}
 	\vspace{-0.7cm}
	\caption{Overview of PrivSketch.}
	\label{fig:overview}
\end{figure*}

\figurename~\ref{fig:overview} provides a high-level overview of PrivSketch workflow. 
At the user end, the encoder encodes items using CMS and the perturber perturbs a sampled one counter in the sketch using RR. 
Then, the perturbed counter $\hat{X}_{k,m}^{(i)}$ is sent to the collector with an ordering matrix $O^{(i)}$ which reflects the order of all counters in the original sketch $X^{(i)}$. 
At the collector end, the decoder restores $\hat{X}_{k,m}^{(i)}$ to the original domain $\mathcal{D}$ by calculating each item's minimum index based on $O^{(i)}$ and updating counts of items $x$ whose minimum index equal to the sampled $k$ and $H_k(x)=m$. 
Then, the calibrator estimates items' frequency by aggregating restored counts from users and calibrating the perturbation error. The protocol is shown in Algorithm~\ref{alg:PrivSketch}. 
We elaborate on its novel designs and details next.

\begin{algorithm}[tb]
{\small
	\SetAlgoLined
	\KwIn{$\{S^{(1)},S^{(2)},\ldots,S^{(n)}\},\epsilon,K,M,D\subset \mathcal{D}$}
	select a set of hash functions $\mathcal{H} = \{H_1, H_2,\ldots, H_K\}$\;%from a pairwise-independent family\; 
	%\tcc*[h]{\footnotesize{Encoder and Perturber}}\\
	\For{each $i \in [1,n]$} 
	{
	    $\hat{X}^{(i)},O^{(i)} \gets$ PrivSketch-User $(S^{(i)},\epsilon,n,K,M,\mathcal{H})$\;
	    send $\hat{X}^{(i)},O^{(i)}$ to the collector\;
	 }
	 %\tcc*[h]{\footnotesize{Decoder and Calibrator}}\\
		\For{each $x \in \mathcal{D}$}
		{
		set $\hat{\mathcal{X}} \gets \{\hat{X}^{(1)}, \hat{X}^{(2)}, \ldots, \hat{X}^{(n)}\} $\;
		set $\mathcal{O} \gets \{O^{(1)}, O^{(2)}, \ldots, O^{(n)}\} $\;
		$\hat{f}(x) \gets $ PrivSketch-Collector
		$(x,\epsilon,n,M,\mathcal{H},\hat{\mathcal{X}},\mathcal{O})$\;
}
	\Return{$\{\hat{f}(x) | x \in D\}$}
} % font size
\caption{PrivSketch}
\label{alg:PrivSketch}
\end{algorithm}

\subsection{Decoding-First Collector-Side Workflow}
An important characteristic of PrivSketch is the decoding-first feature on the collector side, which is designed to reduce the collisions in the private sketching algorithm. 
The naive protocol, PCMS-Min as traditional LDP protocols, consists of three steps: Encode, Perturb, and Aggregate~\cite{Wang2017}. 
Collisions can occur in the Encode and Aggregate procedure.
During encoding, the collision is caused by that different items are hashed into the same positions, which can be reduced by a good choice of the sketching parameters. 
During aggregation, sketches from $n$ users are integrated  into one sketch, equivalent to encoding data from $n$ users using the same sketch. 
This leads to a high probability of collisions due to the large number of users under LDP. 
We find that decoding the perturbed data before aggregation can avoid this collision, where the Decode procedure has been implemented by the collector after the Aggregate procedure but ignored by LDP protocol designers. 
If the collector decodes the perturbed data before aggregation, only the perturbed counts instead of the sketches are aggregated, thus, no collisions.
We present theoretical proof for how the decode-first workflow reduces collision errors following.
Note in our design, we use Calibration instead of Aggregate to describe the procedure where aggregating and calibrating errors caused by the perturbation.

\begin{theorem} \label{theorem:decode-first}
	For estimating the frequency of a value $x\in \mathcal{D}$ using Count-Min Sketch, $\min_k\sum_{i=1}^n X^{(i)}_{k,H_k(x)}$ represents the results of aggregating sketches before decoding, $\sum_{i=1}^n \min_kX^{(i)}_{k,H_k(x)}$ represents the results of decoding sketches before aggregating, the following formula holds:
    {
	\begin{equation}
	\label{eqn:cmp-count-min}
		\min_k\sum_{i=1}^n X^{(i)}_{k,H_k(x)}\ge\sum_{i=1}^n \min_k X^{(i)}_{k,H_k(x)}\ge nf(x)
	\end{equation}
    } % font size
	where $f(x)$ represents the true frequency of $x$. 
\end{theorem}
\begin{proof}
	For each user $U_i$ and any $1\le k\le K$, $X^{(i)}_{k,H_k(x)}$ reflects the occurrence of both $x$ and $x'(x'\neq x)$, which are hashed into the same position with $x$. 
	{
    \begin{displaymath}
		X^{(i)}_{k,H_k(x)}=\mathds{1}\{x\in S^{(i)}\}\vee\mathds{1}\{x'\in S^{(i)},H_k(x)=H_k(x')\}.
	\end{displaymath}
    }
	For the minimum index $k$ where $X^{(i)}_{k,H_k(x)}$ is minimal, the equation above holds. 
	As a result, $\sum_{i=1}^n\min_k X^{(i)}_{k,H_k(x)}  = nf(x) + \sum_{i=1}^n \mathds{1}\{x'\in S^{(i)}, x\notin S^{(i)}, H_{\min_k}(x)=H_{\min_k}(x')\} \ge nf(x)$.
	Moreover, 
%X^{(i)}_{k,H_k(x)}\ge \min_k X^{(i)}_{k,H_k(x)}$ always holds. 
%	Thus, 
%	\begin{displaymath}
		$\sum_{i=1}^n X^{(i)}_{k,H_k(x)}\ge \sum_{i=1}^n\min_k X^{(i)}_{k,H_k(x)}$, $1\le k\le K$.
	%\end{displaymath}
	Considering $\min_k$ is one of the case that belongs to $[1,K]$, we can conclude that $\min_k\sum_{i=1}^n X^{(i)}_{k,H_k(x)}\ge \sum_{i=1}^n\min_k X^{(i)}_{k,H_k(x)}$. 
\end{proof}
\par
Thus, when an unbiased estimation of the query result of the original Count-Min Sketch is achieved, the decode-first collector-side workflow brings fewer errors. 
Next, we introduce how to ensure an unbiased estimation in PrivSketch. 

\subsection{Ordering Matrix Generation}\label{orderingmatrix}

In PrivSketch, the minimum index of the perturbed count can be changed by the randomized response mechanism which hinders an unbiased estimation. 
As shown in \figurename~\ref{fig:overview}, the collector queries the perturbed sketch $\hat{X}^{(i)}$ and estimates based on it. 
Assume the calibration for estimation of the frequency $f(x)$ in $\mathcal{D}$, is based on sketches $\hat{X}^{(i)}$ with a linear function $h(x)$, i.e. $\hat{f}(x)=h(\sum_{i=1}^n \min_k \hat{X}^{(i)}_{k,H_k(x)})$.
PrivSketch needs to satisfy the expectation of the variable after perturbation is an unbiased estimation of the result from querying the original sketch $ \widetilde{f}(x)=\frac{1}{n}\sum_{i=1}^n \min_k X^{(i)}_{k,H_k(x)}$. 
That is, 

{
    \begin{displaymath}
		\mathbb{E}[\hat{f}(x)]  = \mathbb{E}[h(\sum_{i=1}^n \min_k \hat{X}^{(i)}_{k,H_k(x)})] = \widetilde{f}(x) = \frac{1}{n}\sum_{i=1}^n \min_k X^{(i)}_{k,H_k(x)}.
	\end{displaymath}
 }
% \end{small}
Assuming that the row indices of the minimum count for $x$ in the perturbed and original sketches are $k'$ and $k$, 
if $k\neq k'$,
% \begin{small}
{
\begin{align*}
	\mathbb{E}[\hat{X}^{(i)}_{k',H_k'(x)}] =pX^{(i)}_{k',H_k'(x)}-qX^{(i)}_{k',H_k'(x)}=(p-q)X^{(i)}_{k',H_k'(x)} \ge(p-q)\min_k X^{(i)}_{k,H_k(x)},
\end{align*}
}
% \end{small}
where $p$ and $q$ represent the probability of keeping the original value and flipping to the opposite value, respectively.
Due to the randomization, the minimum count in the perturbed sketch is not always in the same position as in the original sketch, i.e., $k\neq k'$. 
However, because the gap between different counts in sketches is diverse and related to the count of specific items, it is difficult to turn the above inequality into an equation by constructing a $h(x)$.
To solve this problem, we propose the ordering matrix.

\begin{figure}[tb]
	\centering
	\includegraphics[width=1\textwidth]{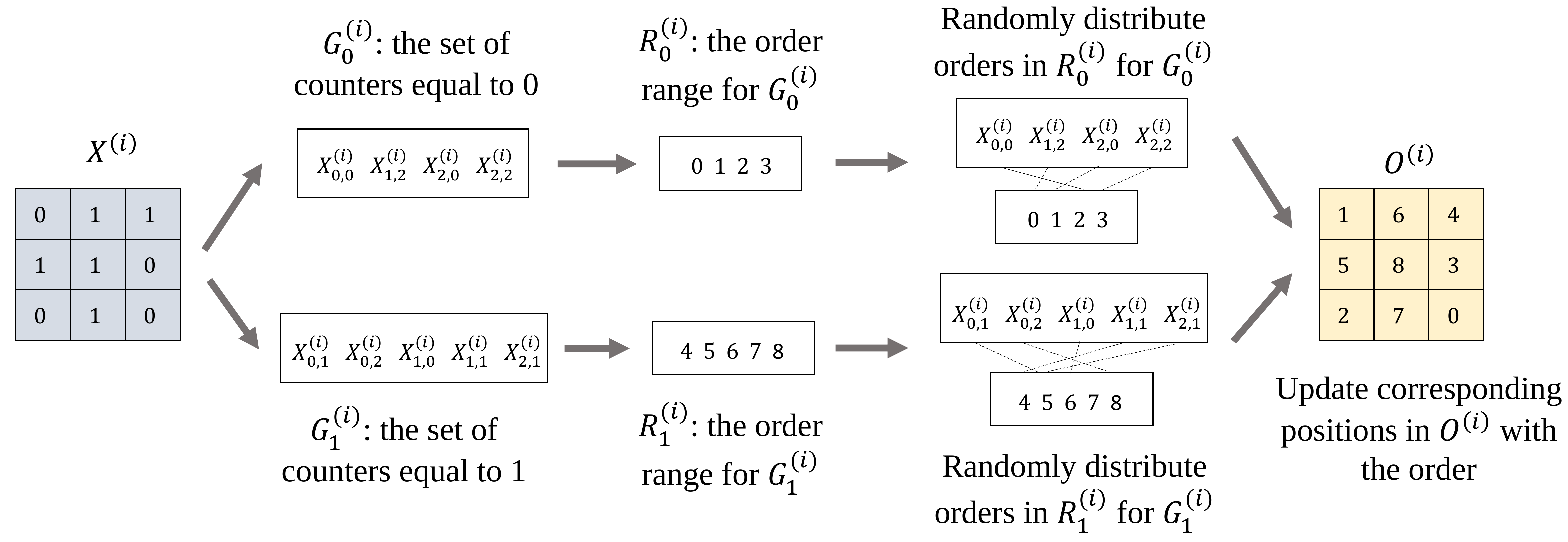}
  \vspace{-0.6cm}
	\caption{The process of generating the ordering matrix.
		\label{fig:ordering-matrix-generation}}
\end{figure}

The ordering matrix $O^{(i)}$ is the background information provided by users, to assist the collector in getting the same row index of the minimum value as the original matrix, which takes advantage of the insensitivity of LDP to any background information to keep the privacy. 
The ordering matrix $O^{(i)}$ is a $K\times M$ matrix, where each position represents the serial number of the corresponding position in the original sketch $X^{(i)}$ ordered by count. 
Firstly, each counter $X^{(i)}_{k,m}$ is distributed into different groups $G^{(i)}_v$ according to its count $v$. 
As a result, $G^{(i)}_v$ includes a set of counters $\{(k,m)|X^{(i)}_{k,m}=v\}$ and its length is denoted by $|G^{(i)}_v|=g_v$. 
Secondly, each group $G_v$ is bound with its order range $R^{(i)}_v = [\sum_{v'\le v}g_{v'}, \sum_{v'\le v}g_{v'}+g_v]$. 
Thirdly, we randomly sample an order without replacement from $R^{(i)}_v$ for each counter in $G^{(i)}_v$ where the order selected for each counter $X^{(i)}_{k,m}$ is denoted as $r^{(i)}_{k,m}$.
Finally, we update the ordering matrix $O^{(i)}_{k,m}=r^{(i)}_{k,m}$. 
Thus, the collector can get the same minimum index by comparing the order of counters in $O^{(i)}$: this has the same result as calculating the minimum index on the original sketch $X^{(i)}$. 
An example is shown in \figurename~\ref{fig:ordering-matrix-generation}.

In the following, 
we prove the estimation is unbiased in PrivSketch (Section~\ref{utility}), and analyze the impact of the ordering matrix on privacy (Section~\ref{PrivacyAnalysis}).

\subsection{Utility Proof and Improvements}\label{utility}

We present the protocol details on the user- and collector-side. 
We prove that the estimations are unbiased, and analyze the variances of errors, then employ sampling to achieve high utility. 

\noindent\textbf{User-side protocol (Algorithm~\ref{alg:user-side-protocol}).} It consists of an encoder (lines 1-4) and a perturber (lines 5-10).
In the encoder, each user records locally whether $x$ appears, because our objective is to obtain the frequency of any value $x$ in $\mathcal{D}$ (instead of counts).
Consequently, an update in the encoder is a boolean disjunction, not an integer addition. 
Each position $X_{k,m}$ is initialized as \emph{False} (i.e., $0$). 
When $x$ is hashed to $X_{k,m}$, the update is $ X_{k,m} =X_{k,m} \vee True=True$ (line 3). 
After encoding, the ordering matrix is computed by the encoder. 
The perturber uses the randomized response mechanism (as in PCMS-Mean) to perturb each value to the opposite value with a probability of $\frac{1}{e^{\epsilon/KM}+1}$ due to at most $K\times M$ different positions.

\begin{algorithm}[tb]
{\footnotesize
	\SetAlgoLined
	\KwIn{$S^{(i)},\epsilon,n,K,M,\mathcal{H}$}
	%\tcc*[h]{\footnotesize{Encoder}}\\
	initialize a sketch $X^{(i)} \gets \{0\}^{K\times M}$\;
	\For{each $\ell \in [1,L^{(i)}]$,each $k \in [1,K]$} 
	{
	   $X^{(i)}_{k,H_k(s_\ell^{(i)})}=1$\;
	 }
	 generate the ordering matrix $O^{(i)}$\;
	 %\tcc*[h]{\footnotesize{Perturber}}\\
		\For{each $k \in [1,K]$, each $m \in [1,M]$}
		{
		    sample $r$ from $[0,1]$ uniformly\;
		    \uIf{$r < \frac{1}{e^{\epsilon/KM}+1}$}
		    {
		    $\hat{X}^{(i)}_{k,H_k(s_\ell^{(i)})}= -2 X^{(i)}_{k,H_k(s_\ell^{(i)})}+1$\;
		    }
		    \Else
		    {
		    $\hat{X}^{(i)}_{k,H_k(s_\ell^{(i)})}=2X^{(i)}_{k,H_k(s_\ell^{(i)})}-1$\;
		    }
        }
        
	\Return{$\hat{X}^{(i)},O^{(i)}$}
}
\caption{PrivSketch-User}
\label{alg:user-side-protocol}
\end{algorithm}

\noindent\textbf{Collector-side Protocol (Algorithm~\ref{alg:collector-side-protocol}).}
First, the decoder estimates the perturbed frequency of the value $x$ using the perturbed minimum in $\hat{\mathcal{X}}$. 
The position of the minimum is provided by the background information $\mathcal{O}$ (line 4). 
Next, the calibrator removes the perturbation error to obtain the final estimation (line 6).
The utility proof of the protocols follows.

\begin{algorithm}[tb]
{\footnotesize
	\SetAlgoLined
	\KwIn{$x,\epsilon,n,M,\mathcal{H},\hat{\mathcal{X}},\mathcal{O}$}
	select a set of hash functions $\mathcal{H} = \{H_1, H_2,\ldots, H_K\}$\;%from a pairwise-independent family\; 
	%\tcc*[h]{\footnotesize{Decoder}}\\
	$C(x) \gets 0$\;
	\For{each $i \in [1,n]$} 
	{
	    $k_{\min} \gets \mathop{\arg\min}\limits_{k}O^{(i)}_{k,H_k(x)}$\;
	    $C(x) \gets C(x) + \hat{X}^{(i)}_{k_{\min},H_{k_{\min}}(x)}$\;
	}
	 %\tcc*[h]{\footnotesize{Calibrator}}\\
	 $\hat{f}(x) \gets \frac{1}{2}(\frac{e^{\epsilon/KM}+1}{e^{\epsilon/KM}-1}\frac{C(x)}{n}+1)$\;
		
	\Return{$\hat{f}(x)$}
}
\caption{PrivSketch-Collector}
\label{alg:collector-side-protocol}
\end{algorithm}

\begin{theorem} \label{theorem:utility}
	Let $C(x)$ denote the perturbed counters for each value in $\mathcal{D}$. 
	$\hat{f}(x)=\frac{1}{2}(\frac{e^{\epsilon/KM}+1}{e^{\epsilon/KM}-1}\frac{C(x)}{n}+1)$ is an unbiased estimation of $\widetilde{f}(x)=\frac{1}{n}\sum_{i=1}^n \min_k X_{k,H_k(x)}^{(i)}$ which is the frequency inferred from the original count-min sketch. 
	Furthermore, the variance of $\hat{f}(x)$ is $\frac{e^{\epsilon/KM}}{n(e^{\epsilon/KM}-1)^2}$.
\end{theorem}

\begin{proof}
	For each user $U_i$, the counters for the item $x$ in row $k$ of perturbed sketch $\hat{X}$ is denoted by $\hat{X}_{k,H_k(x)}^{(i)}$, which value is determined by $X_{k,H_k(x)}^{(i)}$ (lines 6-10 in Algorithm~\ref{alg:user-side-protocol}).
$C(x)$, which represents the result by aggregating the perturbed counters at the minimum position $\min_k\hat{X}_{k,H_k(x)}^{(i)}(x)$, equal to $\sum_{i=1}^n\min_k\hat{X}_{k,H_k(x)}^{(i)}(x)$, satisfies: 
$\mathbb{E}[C(x)] = 2(p\!-\!q)n\widetilde{f}(x)\!+\!(2q\!-\!1)n$,
${\rm Var}[C(x)] = 4n\{(p+q-1)(p-q)\widetilde{f}(x)+q(1-q)\}$,
where $n\widetilde{f}(x)$ is the estimated number of users with $x$ in their sequences using the set of original sketch $\mathcal{X}$. 
	In our protocol, 
	$p=\frac{e^{\epsilon/KM}}{e^{\epsilon/KM}+1}, q=\frac{1}{e^{\epsilon/KM}+1}$.
Thus, the expectation of $\hat{f}(x)$, can be shown to be equal to $\widetilde{f}(x)$ as follows, which means the estimation is unbiased. And its variance of $\hat{f}(x)$ is satisfied:
	{
	\begin{align}
\mathbb{E}[\hat{f}(x)]=\frac{1}{2}(\frac{e^{\epsilon/KM}+1}{e^{\epsilon/KM}-1}\frac{\mathbb{E}[C(x)]}{n}+1) = \widetilde{f}(x)
\label{eqn:prvisketch-expectation}\\ 
		{\rm Var}[\hat{f}(x)] = \frac{1}{4n^2}\frac{(e^{\epsilon/KM}+1)^2}{(e^{\epsilon/KM}-1)^2}{\rm Var}[C(x)] = \frac{e^{\epsilon/KM}}{n(e^{\epsilon/KM}-1)^2}.
        \label{eqn:prvisketch-var}
	\end{align}
	} % font size
	
\end{proof}

\noindent\textbf{Sample the sketches.} Following the above design, larger $K$ and $M$ make the perturbation probability closer to $\frac{1}{2}$ as random.%, which means that the perturbed data are almost random. 
And the variance also increases at the same time. 
The limited privacy budget $\epsilon/KM$ for each counter makes the collector receive scarcely useful information from the perturbed sketches, making it difficult to infer the true frequency. 
To solve the problem, the sampling technique is a common solution, i.e., randomly sampling one from $K\cdot M$ counters on the user end.
Thus, for each counter chosen, the privacy budget becomes $\epsilon$. The variance now is
%as follows:
% {\small
% \begin{equation} \label{eqn:privsketch-var-new}
${\rm Var}[\hat{f}(x)]   = \frac{KM e^{\epsilon}}{n(e^{\epsilon}-1)^2}$,
% \end{equation}
% } 
which is linearly related to $K\cdot M$ due to the sampling error, thus increasing more slowly than the exponential relation in Equation~\eqref{eqn:prvisketch-var}. 
However, it is challenging to obtain the optimal sketching, because as $K$ and $M$ increase, the collision error introduced by Count-Min Sketch decreases, which is also related to the data domain size $d$ and its distribution~\cite{cormode2005improved}.%\footnote{We plan to investigate this in our future work.}. 
Though, we experimentally evaluate the effect of $K$ and $M$ on frequency estimation in Section~\ref{params_experiment}.
Besides, the utility of sampling in sketches is also verified by comparing with traditional PSFO~\cite{WangItemset} in Section~\ref{experiments}.

\subsection{Privacy Analysis} 
\label{PrivacyAnalysis}

When the user sends only the perturbed counter $\hat{X}_{k,m}^{(i)}$ to the collector with the flipping probability $\frac{1}{e^{\epsilon}+1}$, $\epsilon$-LDP is satisfied. 
However in PrivSketch, the user need also send the ordering matrix $O^{(i)}$ to the collector which may expose useful messages and indirectly damage the privacy. 
In the following, we analyze the influence of $O^{(i)}$ on privacy. 

\begin{figure}[tb]
\centering
 \begin{minipage}[t]{1\textwidth}
% \begin{figure}[tb]
	\centering
	\includegraphics[width=1\textwidth]{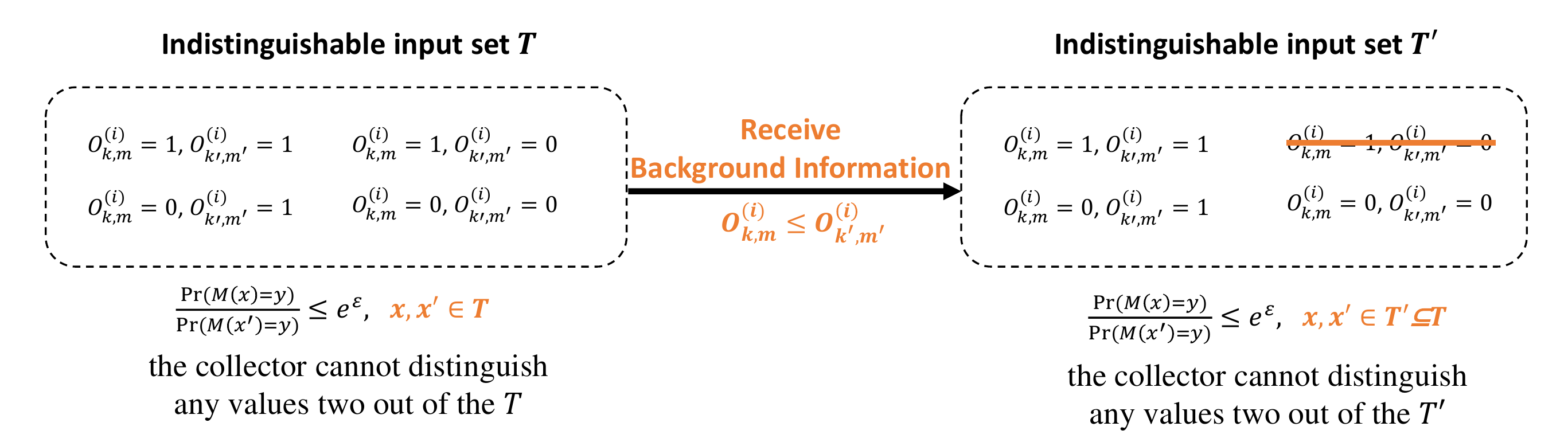}
 	\vspace{-0.7cm}
	\caption{Example of the effect of background information on indistinguishable input set.}
	\label{fig:indisguishable-input-set}
%\end{figure}
\end{minipage}
\end{figure}

\textbf{The ordering matrix $O^{(i)}$ can be utilized to exclude some possible inputs for the collector, but the collector still cannot distinguish some inputs. }
As \figurename~\ref{fig:indisguishable-input-set} shows, if $O^{(i)}_{k,m} \le O^{(i)}_{k',m'}$, $X^{(i)}_{k,m}=1$ and $X^{(i)}_{k',m'}=0$ will not hold at the same time. 
Thus, the cases of the possible sketches of users are reduced from $4$ to $3$ in the collector's view. 
To quantify the effect of the background information, we introduce \textit{indistinguishable input set} to represent the possible inputs in the collector's view,
denoted by $T$. 
According to the $LDP$ definition, any two inputs are indistinguishable regardless of any background knowledge from the adversary. 
Therefore, we can deduce that any two inputs in the indistinguishable set still satisfy the $LDP$ definition, even though the indistinguishable set becomes smaller than without the background information. 

\begin{theorem} \label{theorem:inputset}
	Consider a mechanism $\mathcal{M}$ that satisfies $\epsilon$-LDP, its indistinguishable input set $T$, and any two inputs $x, x'$.
	When the collector receives any output $y$, along with the background information $I$, there exists an indistinguishable set $T'\subseteq T$ satisfying the following inequality:
	% {\small
	% \begin{equation}
		$\frac{\Pr[M(x)=y]}{\Pr[M(x')=y]}\le e^{\epsilon},\quad x,x' \in T'$.
	% \end{equation}
	% } % font size
\end{theorem}
\begin{proof}
    For any $I$, $T$ can be divided into two parts, $T_+$ and $T_-$. 
    The former represents the inputs that are consistent with the information $I$, i.e., the possible inputs when $I$ is true. 
    The latter includes the inputs that contradict the information $I$, that is, the impossible inputs when $I$ is true. 
    Based on $I$, the collector can infer that the original input belongs to $T_+ (\subseteq T)$.  
    For any two inputs $x,x'\in T_+$, $x,x'$ is also in $T$. 
    Therefore, following the definition of $\epsilon$-LDP, $\frac{\Pr[M(x)=y]}{\Pr[M(x')=y]}\le e^{\epsilon}$ is satisfied and any two input $x,x' \in T'$ is distinguishable.
\end{proof}

\textbf{The indistinguishable input set $T'$ computed by the ordering matrix $\mathcal{O}$ is enough to protect the privacy of users in our problem. }
In PrivSketch, what each user needs to protect is its original sketch matrix $X^{(i)}$. 
Thus, the collector should not infer the value of any counter in $X^{(i)}$ is $1$ or $0$.
In PrivSketch, counters can be divided into two groups, $G_1$ and $G_{0}$, and $g_1+g_{0}=KM$. 
Thus, when the collector receives $O^{(i)}$, the indistinguishable input set $T'$ at most includes $KM+1$ possible sketches with different sizes of each group. 
There are some constraints for sketches, e.g., it is impossible that $g_1 = 1,2,3$, because when there is an item occurred, for each $k\in[1,K]$, $\exists (k,m)\in G_{1}, m\in[1,M]$. 
Nevertheless, $\{0\}^{KM},\{1\}^{KM}\in T'$ always holds. 
Thus, there is no counter with same value in different possible inputs, that is, its value is equal to $1$ in some inputs and equal to $0$ in the other inputs. 
The collector still cannot determine the value of each counter, which is sufficient to protect the privacy of users.

\section{Experimental Evaluation}

In this section, we evaluate the utility and running time of PrivSketch over synthetic and real datasets, and analyze how the main parameters affect its performance. 
For a comprehensive evaluation, we compare PrivSketch to the state-of-the-art PCMS-Mean~\cite{Apple2017}, and PSFO~\cite{WangItemset} based on OLH~\cite{Wang2017} (denoted as PS-OLH in our experiments) for frequency estimation, and SVIM~\cite{WangItemset}, a two-phase heavy hitter discovery protocol for discovery of frequent items. 

\noindent\textbf{Environment.} 
We implement all LDP protocols in Python and conduct experiments on a server with 2 Intel Xeon 3206R Processors and 32G RAM running Centos. 
We repeat each experiment 10 times and report the average results.

\noindent\textbf{Datasets.} 
We use 6 synthetic datasets and 2 real datasets (see Table\ref{tab:datasets}).

\noindent$\bullet$
\textbf{Synthetic Datasets}: These datasets follow Zipf distribution that real data stream often conforms to, with different number of users $n$ and domain size $d$.

\noindent$\bullet$
\textbf{Kosarak}~\cite{Kosarak}: This dataset contains the clicked items that anonymized users from a Hungarian online news portal, involving nearly 1M users and 40K items. 
	
%	\item 
\noindent$\bullet$
\textbf{AOL}~\cite{AOL}: This dataset contains search queries of users on AOL between March 1 and May 31, 2016, with corresponding URLs clicked by them. 
The dataset includes more than 500K users with 1.6 million distinct URLs. 
%\end{itemize}

	\begin{table}[tb]
		\caption{Datasets characteristics}
            \centering
		\label{tab:datasets}
			\begin{tabular}{c c c c c c||c c c c c c}
				\hline
				\bfseries Dataset & \bfseries $n$ & \bfseries $d$ & \bfseries $\max$ & \bfseries $\min$ & \bfseries $P_{90}$ & \bfseries Dataset & \bfseries $n$ & \bfseries $d$ & \bfseries $\max$ & \bfseries $\min$ & \bfseries $P_{90}$\\
				\hline
				Kosarak & 990002 & 41270 & 2498 & 1 & 15 & AOL & 521693 & 1632788 & 61932 & 1 & 62 \\
				% \hline
				Dataset1 & 100000 & 100000 & 123 & 1 & 80 &         Dataset2 & 10000 & 100000 & 117 & 1 & 78\\
    				% \hline
				Dataset3 & 100000 & 20000 & 112 & 1 & 73 & Dataset4 & 100000 & 40000 & 107 & 1 & 72\\
				% \hline
				Dataset5 & 100000 & 60000 & 110 & 1 & 74 & Dataset6 & 100000 & 80000 & 109 & 1 & 75\\
				\hline
			\end{tabular}
        %\vspace{-0.2cm}
	\end{table}

\noindent\textbf{Parameters.}
The number of hash functions $K$ is set to $4$, and each hash function's hash domain size $M$ is set to $128$. 
The default privacy budget $\epsilon$ is $3$, within the acceptable range in many works~\cite{PureLDP,PEM,LDPMiner}. 

\noindent\textbf{Evaluation Measures.} 
We use the following measures, including running time.

\noindent$\bullet$
\textbf{Mean Squared Error (MSE)}.
    We evaluate the frequency estimation accuracy by MSE: $\frac{1}{d}\sum_{x\in\mathcal{D}}(\hat{f}(x)-f(x))^2$,
    where $f(x)$ is $x$'s true frequency. 

\noindent$\bullet$
\textbf{Variance (Var)}. 
	We measure the error of estimating the top-k frequency terms using variance: %, to more comprehensively present our protocol: 
 $\frac{1}{|C_e\cap C_t|}\sum_{x\in C_e\cap C_t}(n\hat{f}(x)-nf(x))^2$.
%	\item 

\noindent$\bullet$
\textbf{Normalized Cumulative Rank (NCR)}. 
To evaluate the estimation of frequent items, NCR measures how many top-k items are identified by the protocol with a quality function $q(.)$. 
It is calculated as follows: $\sum_{x\in C_e}q(x)/\sum_{x'\in C_t}q(x')$,
	where $C_t$ and $C_e$ represents the true top-k items and the estimated top-k items respectively. For $x \in C_t$ with a rank $i$, $q(x)=k+1-i$. For $x \notin C_t$, $q(x)=0$.

\subsection{Comparing to Advanced Protocols} 
\label{experiments}

\begin{figure}[tb]
		\centering
   % \vspace{-0.4cm}
   \hspace{-0.6cm}
		\includegraphics[width=0.29\textwidth]{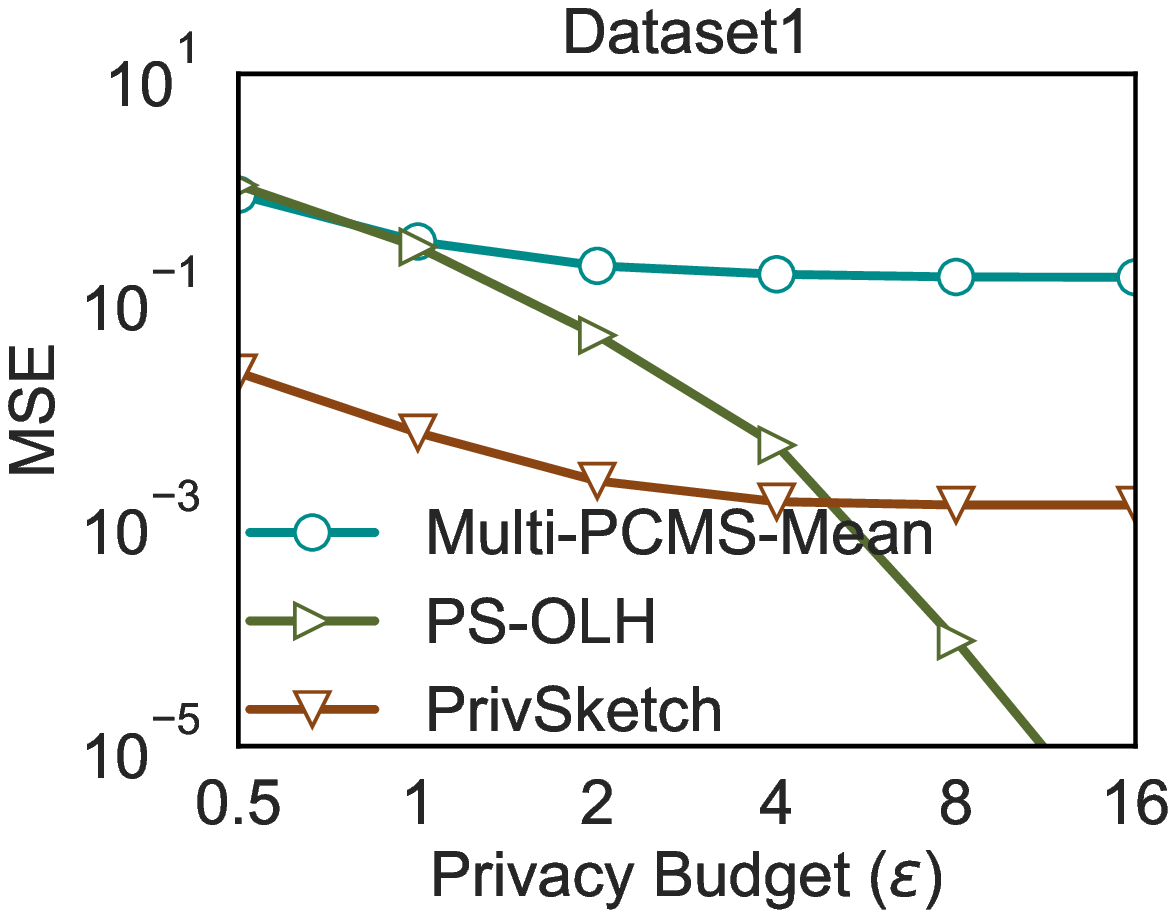}
        \hspace{0.3cm}
   	\includegraphics[width=0.29\textwidth]   {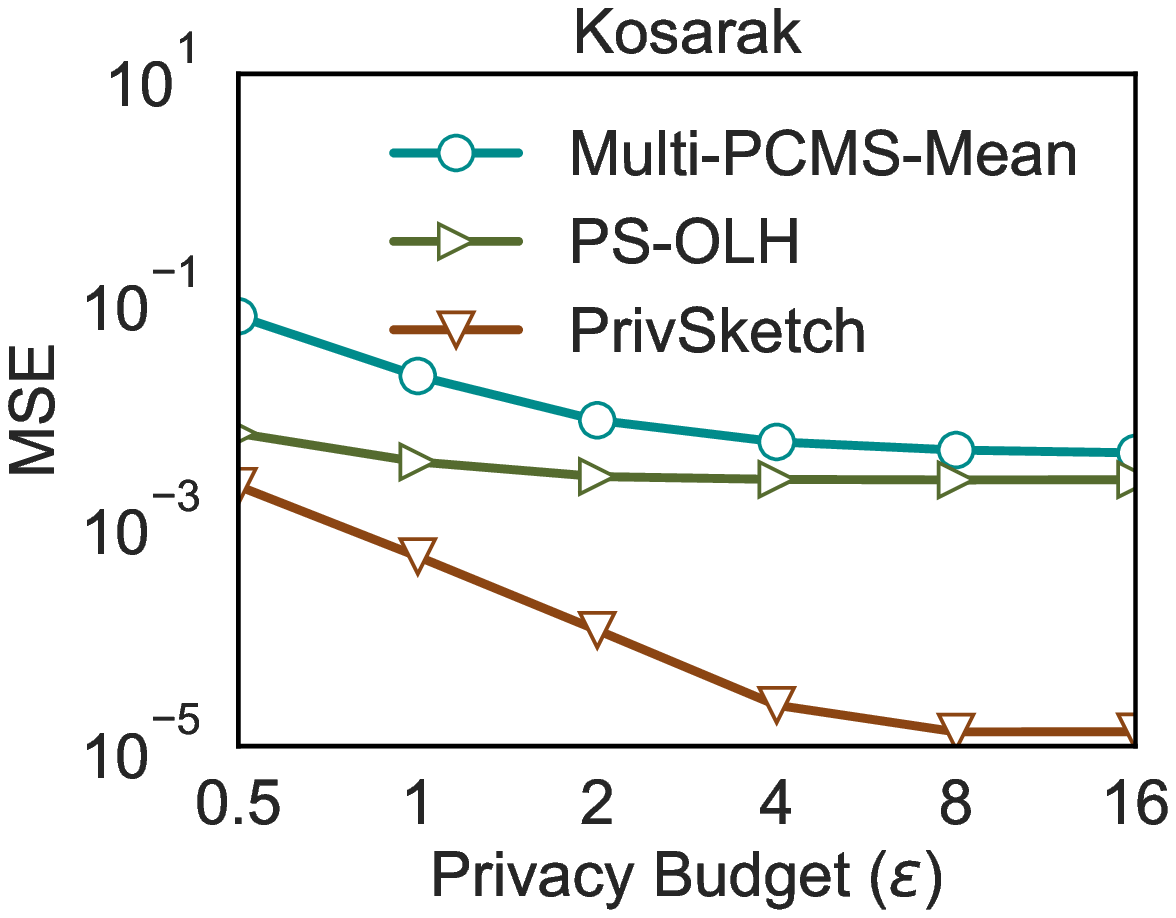}
    \hspace{0.2cm}
        \includegraphics[width=0.29\textwidth]{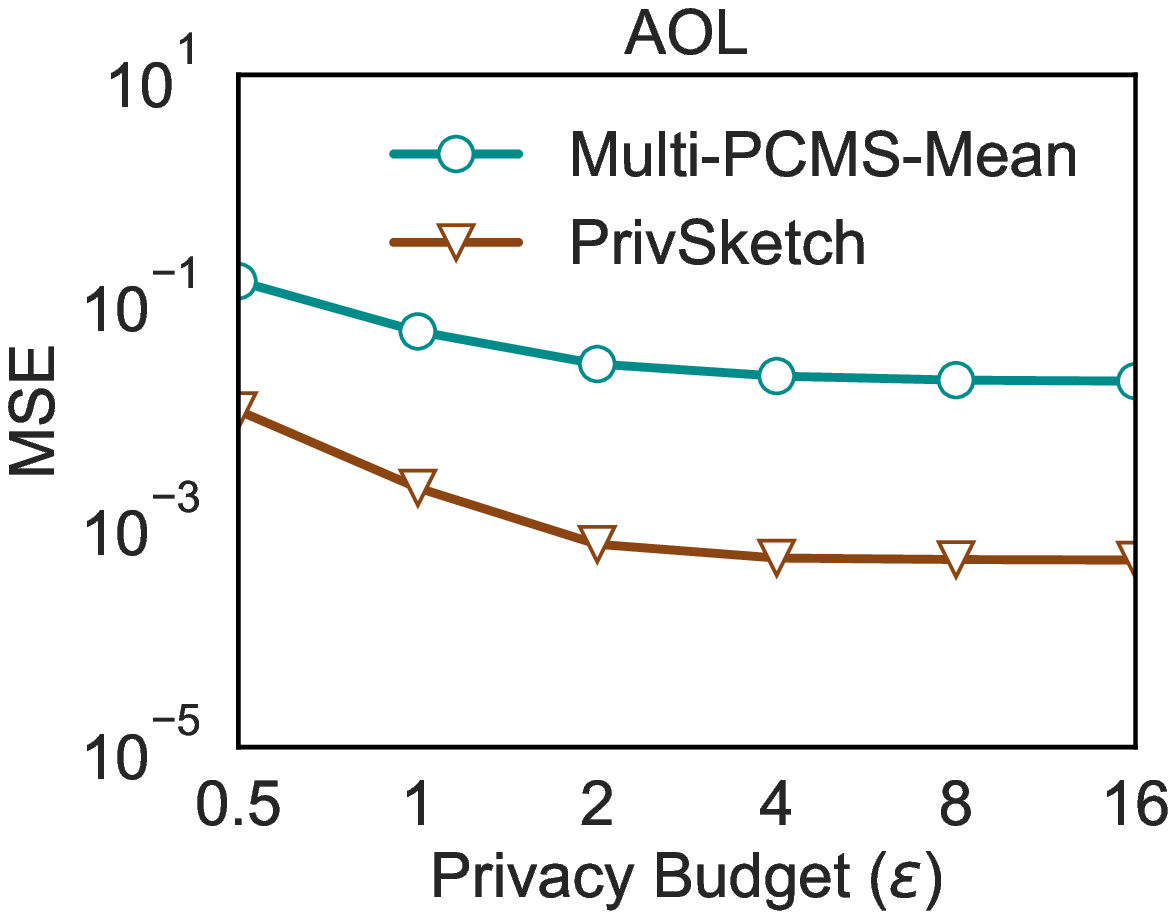}
 \vspace{-0.4cm}
\caption{Experimental results for frequency estimations.} 
\label{fig:exp-freq} 
% \vspace*{-0.58em}
\end{figure}

\noindent\textbf{Experiments for Frequency Estimation:} We compare our protocol to two advanced solutions: (\romannumeral1) a sketch-based solution, Multi-PCMS-Mean, which is an extended version of PCMS-Mean~\cite{Apple2017} for multi-item collection, 
and (\romannumeral2) a non-sketch-based solution, PS-OLH, which is an advanced PSFO~\cite{WangItemset}.
%suitable for multi-item collection. 
PSFO~\cite{WangItemset} combines the padding and sampling technique with a basic frequency estimation protocol to transform multiple-item into one-item problems. 
Because the optimal local hash (OLH)~\cite{WangItemset} performs best when $d\ge 3e^\epsilon+2$ (i.e., for large domains), we choose the PSFO with OLH, i.e., PS-OLH, as our competitor. 
For a fair comparison, %the privacy budget is only used to estimate frequency. 
%Thus, 
we assume the distribution of user input length is known and set the padding length $l$ of PS-OLH to the 90th percentile of the user input~\cite{LDPMiner} (avoiding to use the privacy budget to estimate $l$).

We evaluate the MSE of frequency estimation under different privacy budgets, varying from $0.5$ to $16$, on synthetic and real datasets.
As shown in \figurename~\ref{fig:exp-freq}, PrivSketch performs best, especially for small privacy budgets, which indicates the high utility of PrivSketch and its strong privacy protection.

\begin{figure}[tb]
	\begin{minipage}[t]{\columnwidth}
		\centering
		\includegraphics[width=0.32\textwidth]{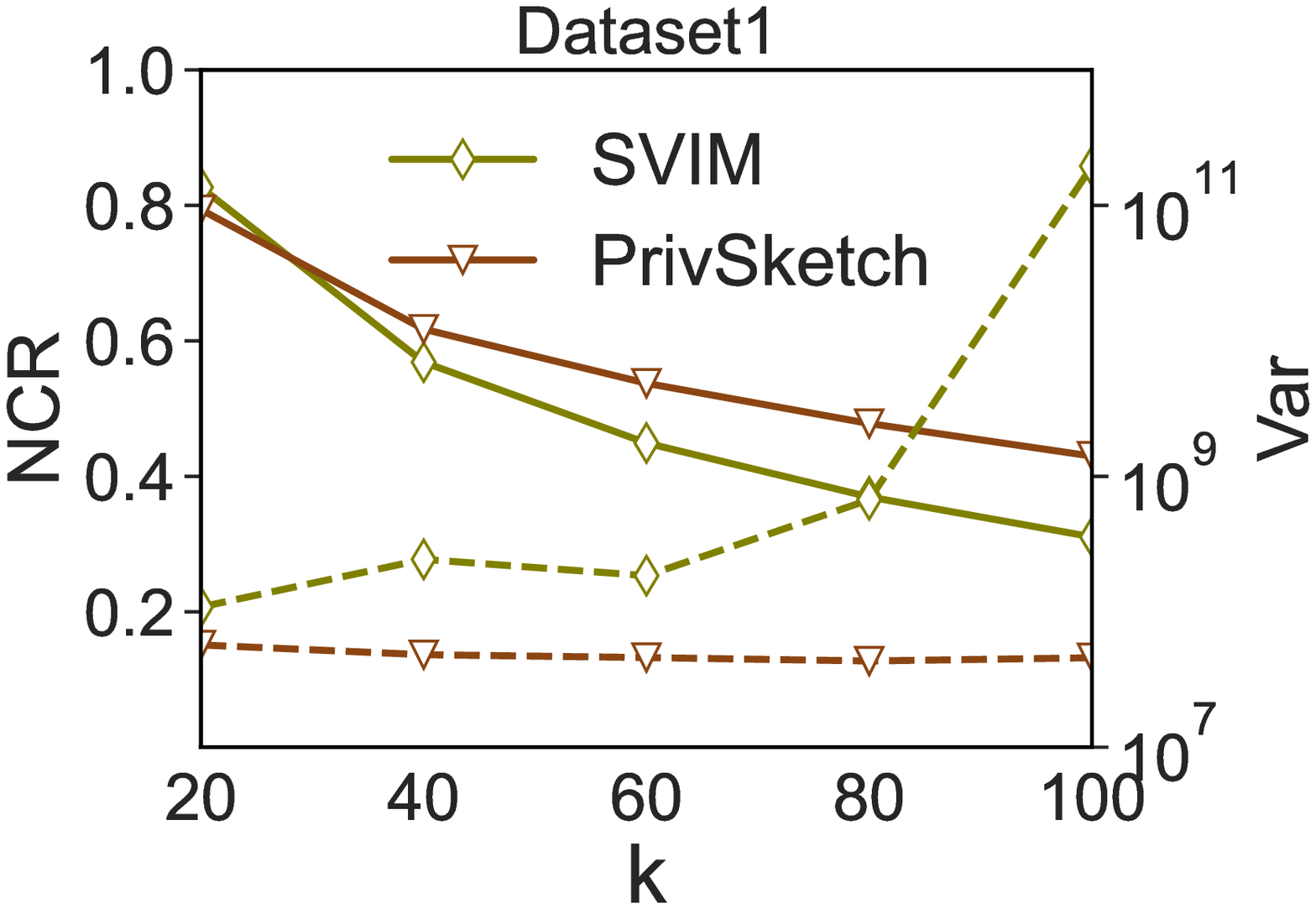}
  		\includegraphics[width=0.32\textwidth]{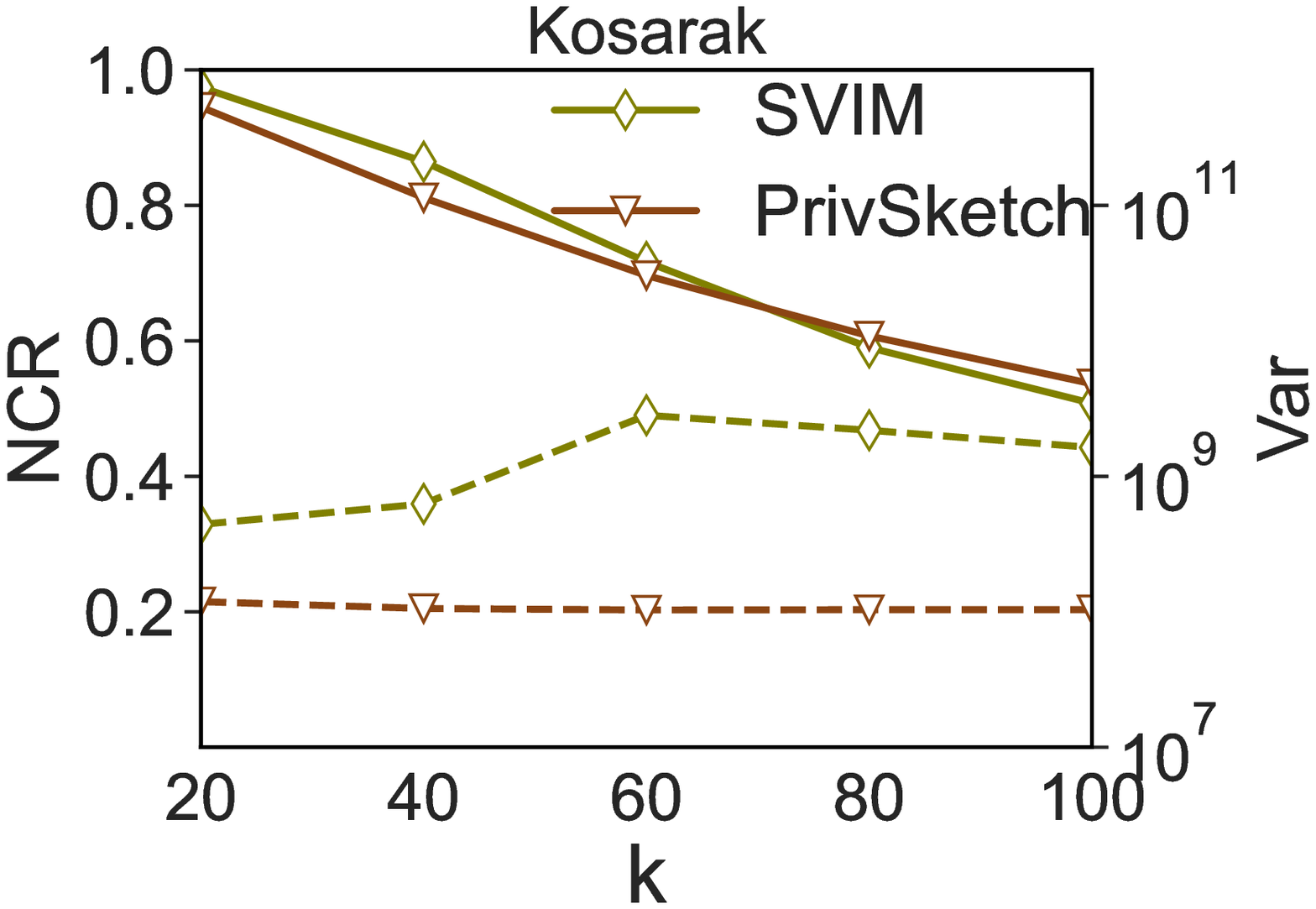}		\includegraphics[width=0.32\textwidth]{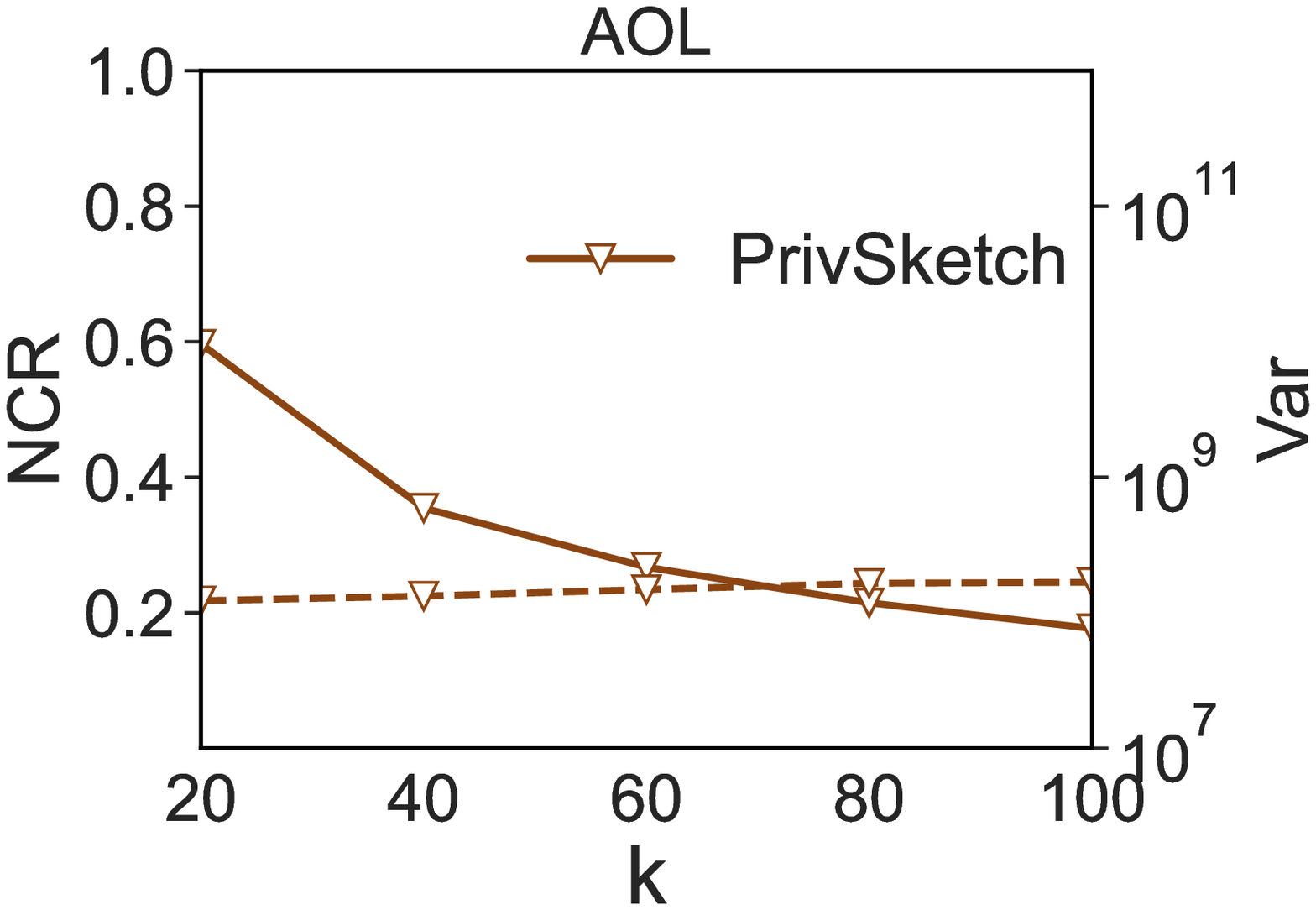}
	\end{minipage}
	% }
 \vspace{-0.8cm}
        \caption{VAR and NCR when varying parameter $k$.} 
         \vspace{-0.2cm}
	\label{fig:exp-topk} 
% \vspace*{-1.58em}
\end{figure}

\noindent\textbf{Experiments for Frequency Item Estimations:} We also evaluate the performance of PrivSketch in frequent item mining (i.e., heavy hitter discovery), a popular application of frequency estimation. 
We compare it with the existing advanced multi-phase protocol, SVIM~\cite{WangItemset}, which is the improved work after LDPMiner~\cite{LDPMiner} and is also applicable in large domains. 
As shown in \figurename~\ref{fig:exp-topk}, PrivSketch performs better than SVIM, especially in frequency estimation for top-k items.
It is expectable because PrivSketch has been designed for accurate frequency estimation, not frequent item identification.

\begin{figure}[tb]
	\centering
     \subfigure[Frequency estimation] 
	{\includegraphics[width=0.315\columnwidth]{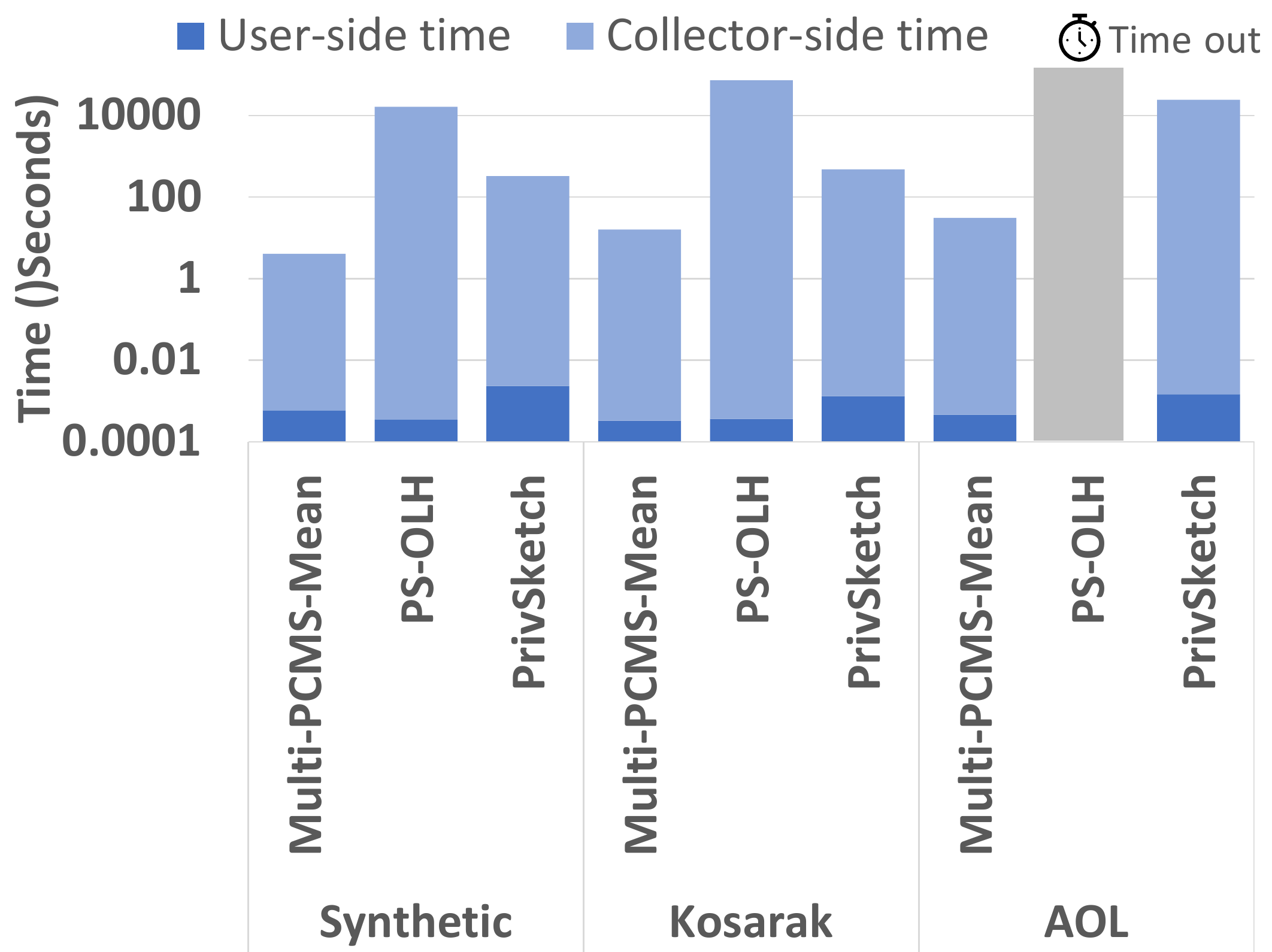}
		\label{fig:exp-all-time}} 
	\subfigure[Frequent item estimation] 
	{\includegraphics[width=0.323\columnwidth]{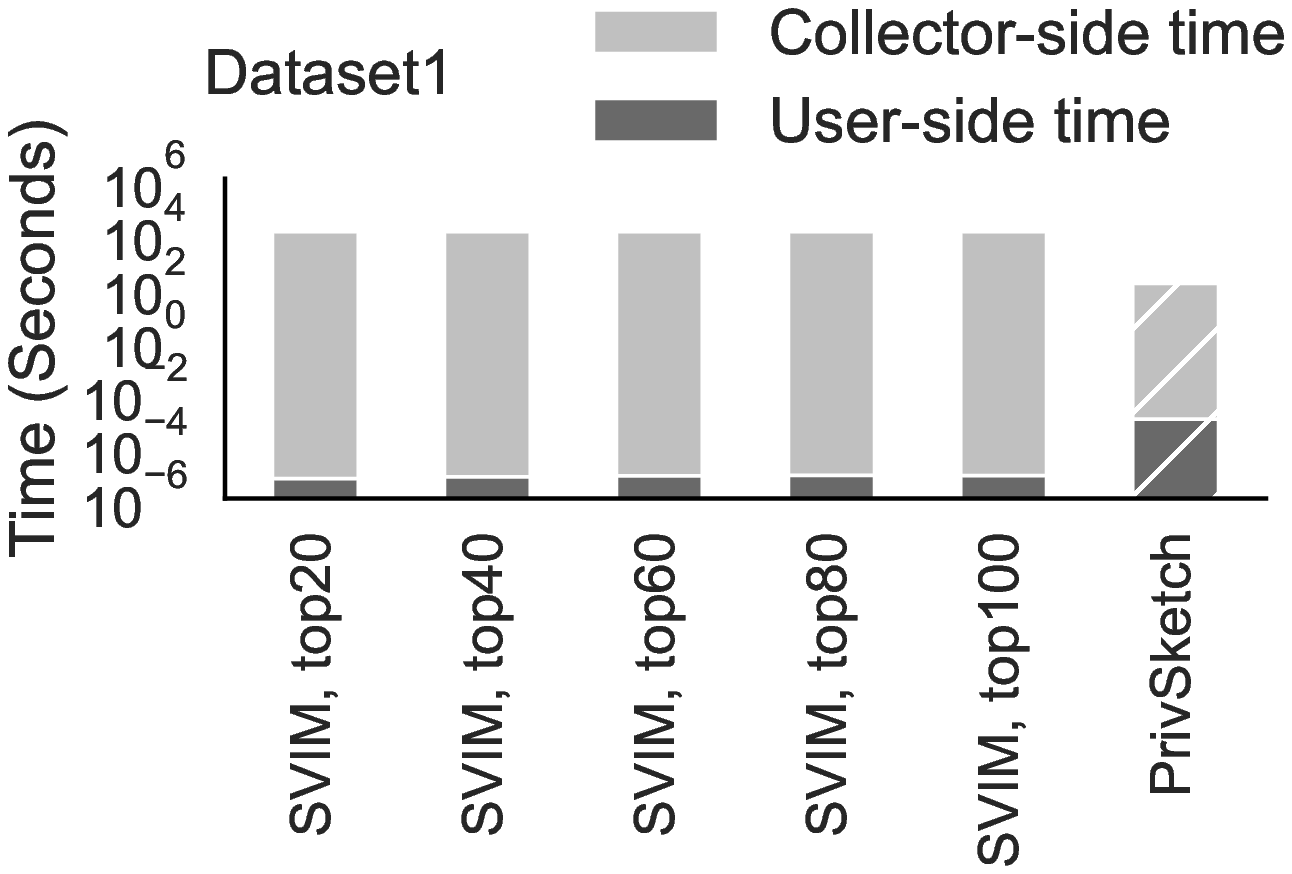}
        \includegraphics[width=0.323\columnwidth]{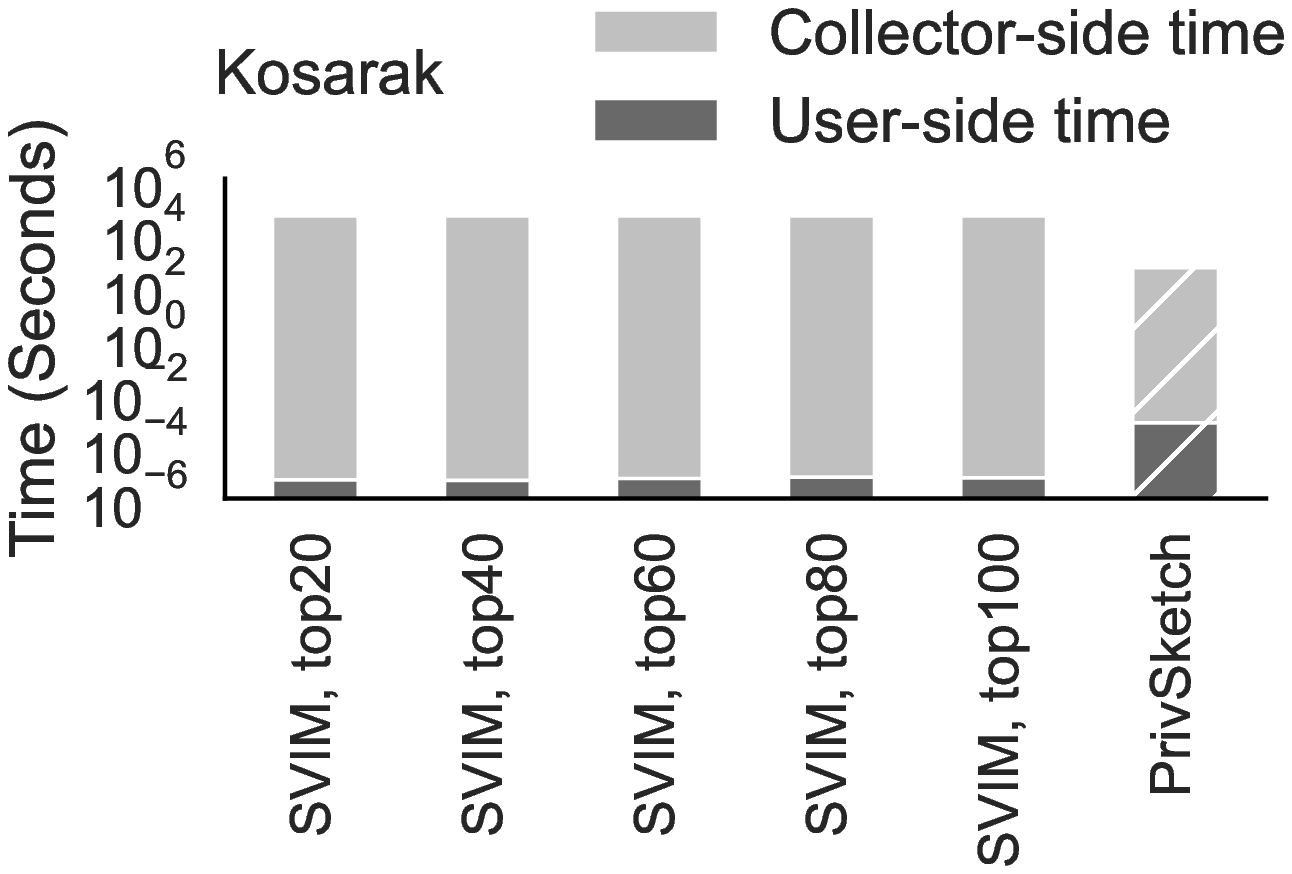}	
		\label{fig:exp-svim-time}} 
	  \vspace{-0.4cm}
\caption{Comparison of running times.}\label{fig:exp-time}
 % \vspace{-0.4cm}
\end{figure}

\noindent\textbf{Evaluation of running time.}\label{exp-time}
As shown in \figurename~\ref{fig:exp-time}, Privsketch maintains a user-side running time smaller than $0.01s$ while performing the calculation of the ordering matrix. 
Overall, PrivSketch is faster than PS-OLH and SVIM about 100 times, but slower than Multi-PCMS-Mean with much larger MSEs (in \figurename~\ref{fig:exp-freq}).
The long running time of PS-OLH, SVIM and PrivSketch is the sacrificed time of reducing domain cardinality to gain high utility.
Thus they need to restore the estimated items to the original domain for each user on the collector side, resulting in a complexity of $\mathcal{O}(nd)$. 
In PrivSketch, each user shares the same hash functions instead of local hash functions used in PS-OLH and SVIM, resulting in fewer hash function calculations on the collector side. 
We omit experimental results for PS-OLH and SVIM over AOL, because they need more than 10 days to compute, making them cumbersome to use in practice. 

\subsection{Experiments with Different Parameters} \label{params_experiment}

In this section, we compare PrivSketch with other sketch-based solutions to present the effect of our design under different parameters.
In addition to Multi-PCMS-Mean, its min-estimation variant (denoted by Multi-PCMS-Min) and a middle version of PrivSketch without sampling (denoted by PrivSketch-noSmp) are also compared, to show the better utility of min estimation and the effect of our decode-first and sampling design.

\begin{figure*}[tb]
% \vspace*{0.2em}
\begin{minipage}[b]{0.24\columnwidth}
 %\vspace*{0.2cm}
   	\includegraphics[width=\textwidth]   {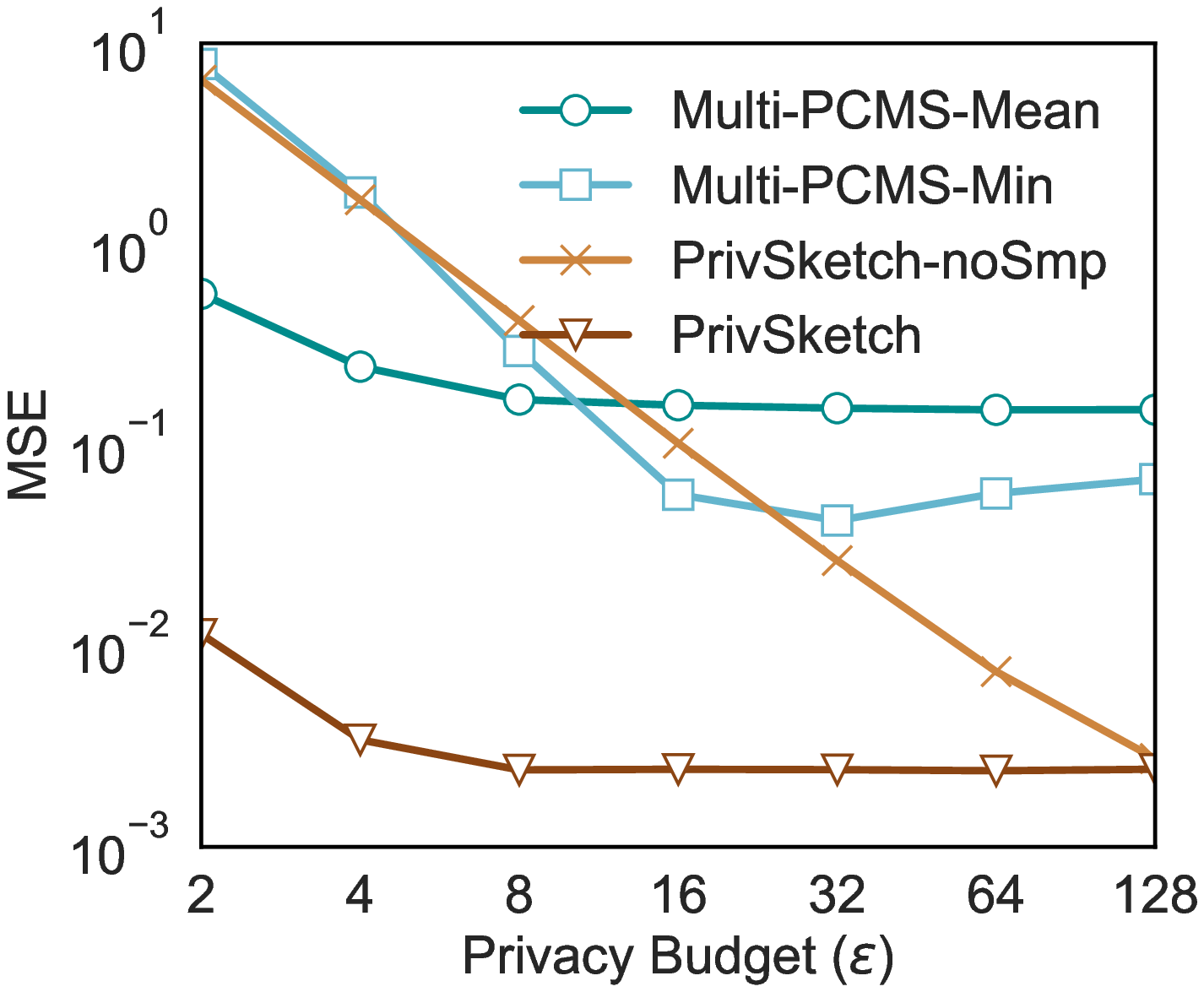}
\vspace*{-0.7cm}
\caption{MSE when varying $\epsilon$, $n=10^4$.}
\label{fig:exp-vary-n} 
\end{minipage}
\hspace{0.47cm}
\begin{minipage}[b]{0.76\columnwidth}
\vspace*{-0.8cm}
        \centering
	\subfigure[vary M (K=4)] 
	{\includegraphics[width=0.32\columnwidth]{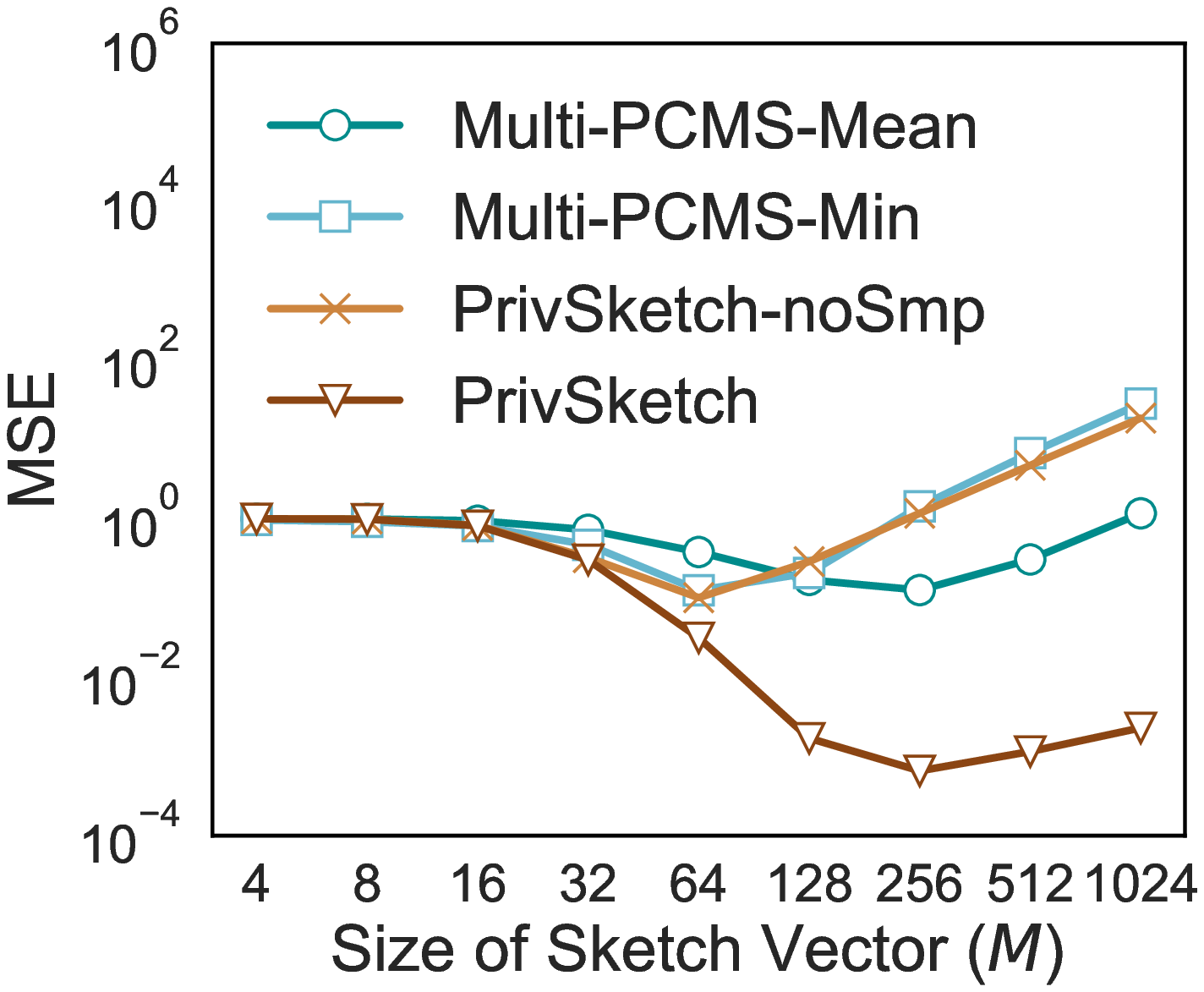}
		\label{fig:exp-syn-KM-K4}} 
	\subfigure[vary K (M=32)]
	{\includegraphics[width=0.32\columnwidth]{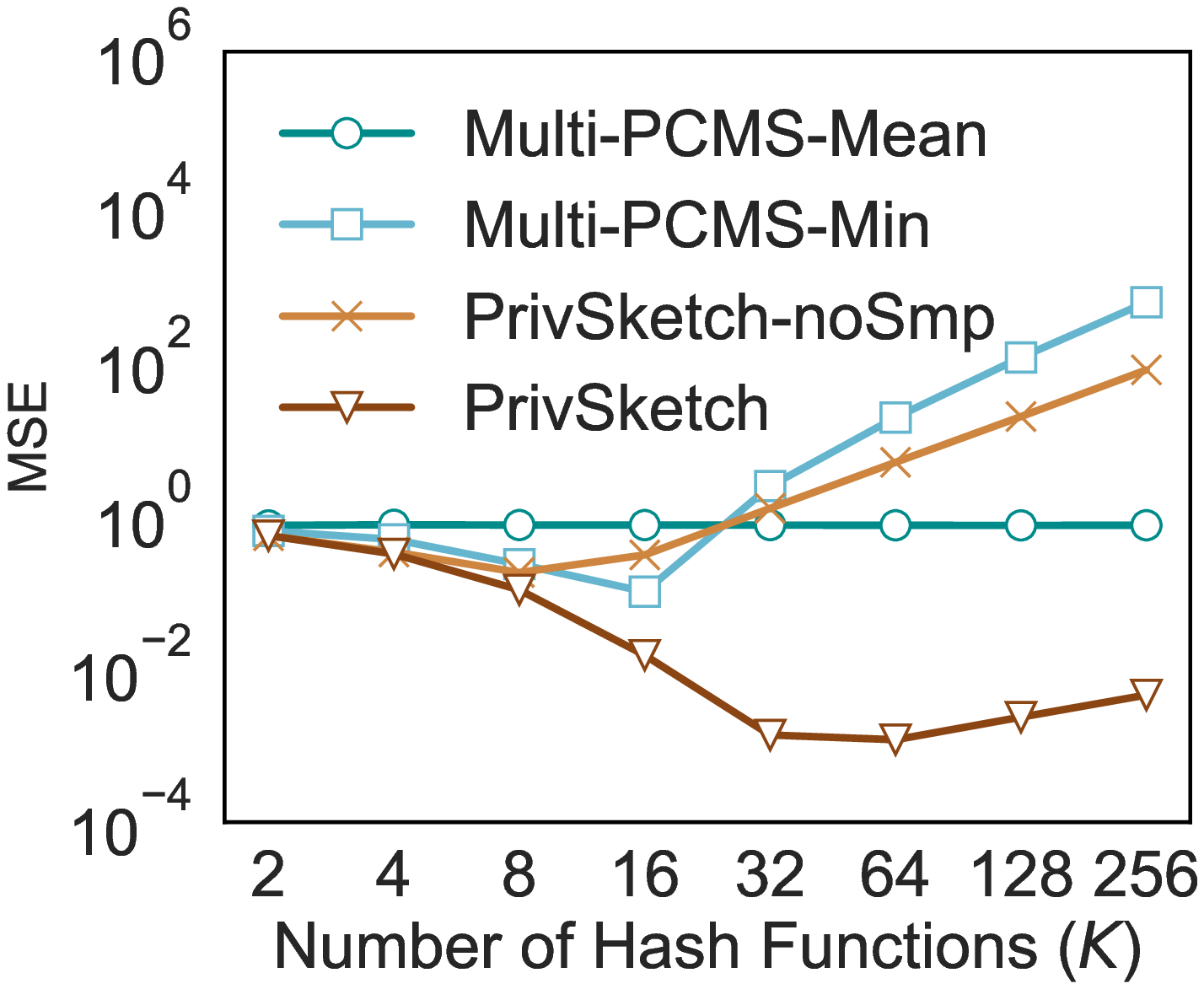}
		\label{fig:exp-syn-KM-M32}}
	\subfigure[vary d]
	{\includegraphics[width=0.3\columnwidth]{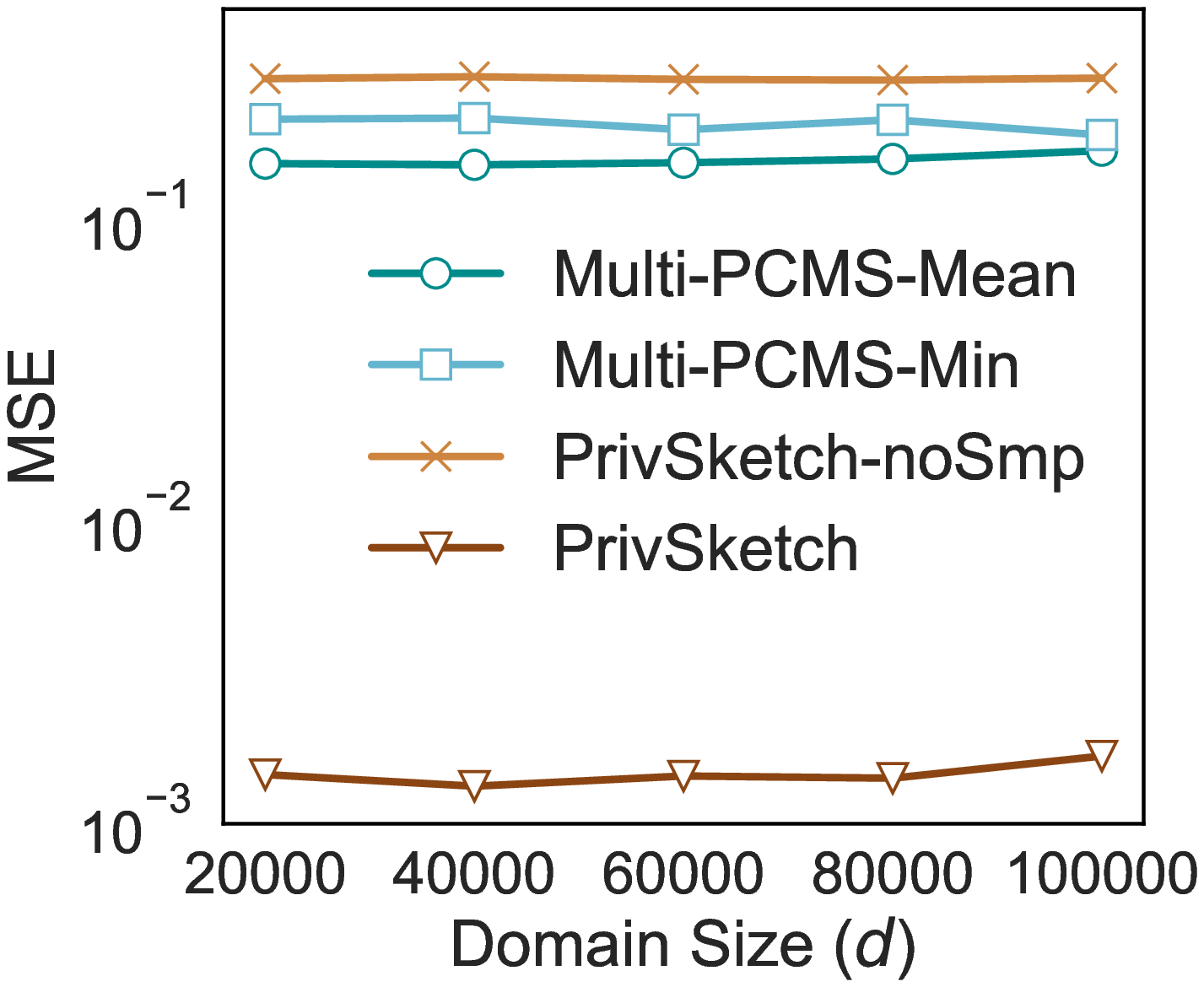}
		\label{fig:exp-syn-d}}
    \vspace{-0.47cm}
	\caption{MSE when varying parameters $K$, $M$, $d$.} 
    \vspace{-0.2cm}
	\label{fig:exp-syn-KM} 
\end{minipage}
% \vspace*{-1.58em}
\end{figure*}

\noindent\textbf{Utility with small number of users.}
We evaluate the MSE on Dataset2 with $10^4$ users under a privacy budget range $[2,128]$. 
Note the unrealistic privacy budget used here is to show the effect of our designs.
In \figurename~\ref{fig:exp-vary-n}, PrivSketch always performs best especially under a small $\epsilon$. 
We observe similar results (omitted for brevity) when $n$ varies in $[10^4$,$10^6]$. 
The result verifies that decode-first workflow with the ordering matrix effectively reduces the collision probability of sketches, and the min estimation has better accuracy than the mean estimation.

\noindent\textbf{Impact of the size of the domain.} 
We conduct this experiment on a group of synthetic datasets, which sets $K, M$ and $\epsilon$ with default values, fixes $n = 10^5$, and varies the domain size $d$. 
As shown in \figurename~\ref{fig:exp-syn-d}, the errors of the four protocols only slightly increase with the increase of $d$. 
Theoretically, in a larger domain, when the sketch size is fixed, the collision probability increases, leading to an increase in error. 
However, since the items held by each user are sparse compared to the domain space, and the distribution of the number of items held by each user changes a little, the domain size change has a small impact. 
This confirms that sketching is an effective domain reduction and encoding method for data collection from a large domain.

\noindent\textbf{Impact of the parameters of the sketch.}
In \figurename~\ref{fig:exp-syn-KM}, we evaluate the effects of different $K$ and $M$ of the sketching using the datasets with parameters $n = 10^5$, $d = 10^5$ under $\epsilon=3$. 
In \figurename~\ref{fig:exp-syn-KM-K4}, we can see that the utility of the PrivSketch is far better than the other three protocols under different $M$ while fixing the hash vector size $K=4$. 
As expected, increasing the size of the hash vector can reduce the estimation error.
However, when $M$ increases to a specific value, the error does not decrease but increases. 
This is because $M$ affects two types of errors in these sketch-based LDP protocols. 
When $M$ increases, the collision probability decreases, but the perturbation probability or the sampling errors increases.
Varying the $K$ brings a similar result to $M$, as shown in \figurename~\ref{fig:exp-syn-KM-M32}.
However, the effects of $K$ on Multi-PCMS-Mean protocol is different. 
Changes of $K$ do not affect its MSE, 
because the effect of choosing one of the $K$ hash functions when encoding is eliminated by the sum of $K$ counters corresponding to $K$ hash functions during the estimation process.

\section{Related Work}

\textbf{Set-valued Data Collection.} The diverse set size is a challenge for set-value data collection under LDP.
Padding and Sampling~\cite{LDPMiner} is a common way to unify the set length, such as in PSFO~\cite{WangItemset}, PrivSet~\cite{ShaoweiWang2018}.
Although Wang~\cite{Wang2020} proposes the wheel mechanism to reduce the computational overhead, 
these works do not aim at a large domain, where an efficient data structure is needed. 
Many works~\cite{LDPMiner,WangItemset,TreeHist,PEM},
focus on frequent item mining, also known as heavy-hitters discovery in a huge domain. 
They utilize a multi-phase strategy to reduce the large domain size first, using a small part of the privacy budget to discover frequent candidates, and using the remaining part to obtain an accurate estimation. 
Nevertheless, this strategy is not suitable for estimating frequency.

\noindent \textbf{Frequency estimation with Hash-Encoding Technique.}
Under LDP settings, to reduce the data domain, RAPPOR~\cite{RAPPOR} adopts Bloom filters to encode data, which requires expensive computations to use LASSO regression for the estimation. 
OLH~\cite{Wang2017} utilizes local hash functions to encode the user data, which requires a large number of hash calculations.
With a simpler estimation solution, Count-Mean Sketch~\cite{Apple2017} was proposed to compute the populated emoji in IOS. 
\cite{vepakomma2021dams,Piao2021} improve it by sending multiple sketches for each user, which also brings extra communication costs. 
\cite{TreeHist} uses Count-Median Sketch with the Hadamard transform when computing the heavy hitters. 
\cite{PureLDP} analyzes and compares LDP protocols with different sketching algorithms, including the Count-Min Sketch. 
However, these protocols, designed for the one-item collection, do not consider the error introduced by the sketching algorithm. 
Recently, \cite{zhou2022locally} utilized hash functions to compute the frequency and the mean estimation of the $k$-sparse vector, 
%which gives upper and lower bounds for this problem 
with an assumption on the number of items each user generates.

\noindent \textbf{Variants of LDP.} 
Lots of works focus on optimizing the variants of LDP to improve its utility. 
Some works introduce extra trust in LDP, such as shuffling anonymized reports from users~\cite{cheu2019distributed,erlingsson2019amplification}, and combining the centralized DP with the local version~\cite{avent2017blender}. 
Some works introduce an extra parameter to relax the privacy constraint, 
such as~\cite{alvim2018metric,gursoy2019secure} that use the distance metric of two inputs to improve the utility,
which is inspired by the geo-indistinguishability concept~\cite{chatzikokolakis2013broadening}. 
Finally, some studies propose discriminative LDP based on different aspects, such as personalized privacy demand~\cite{chen2016private,yiwen2018utility}. %, and distinct sensitivity of the input data~\cite{murakami2019utility,gu2020providing}. 
These works do not utilize the background information to enhance utility as we do in this paper.

\section{Conclusions}

This paper studies the frequency estimation problem %in network traffic data collection 
under local differential privacy. 
We propose a privacy-preserving %network traffic 
data collection protocol, PrivSketch, 
which does not expose the original value of any counter in the sketch.
We experimentally verify the %utility and 
effectiveness of PrivSketch: %and the effectiveness of each design in PrivSketch through experiments, 
it outperforms existing LDP protocols by 1-3 orders of magnitude and executes up to $\sim$100x faster. 

\vspace*{0.40cm}\noindent{\bf Acknowledgments}
We sincerely thank Dr. Zhenyu Liao for his insightful and constructive comments and suggestions on mathematical proof that help to improve the quality. This work is funded by NSFC Grant No. 62202450, Huawei New IP open identification resolution system project No. TC20201119008 and Postdoctoral Exchange Program No. YJ20210185.

\bibliographystyle{splncs04}
\bibliography{ref}

\end{document}